\documentclass{article}
\title{Decouple-and-Sample: Protecting sensitive information in task agnostic data release}

\usepackage[font=small,labelfont=bf]{caption}
\usepackage{tikz}
\usepackage{overpic}
\usepackage{wrapfig}
\usepackage{enumitem} 
\usepackage{overpic} 
\usepackage{color}
\usepackage{amsmath,amssymb}

\definecolor{turquoise}{cmyk}{0.65,0,0.1,0.3}
\definecolor{purple}{rgb}{0.65,0,0.65}
\definecolor{dark_green}{rgb}{0, 0.5, 0}
\definecolor{orange}{rgb}{0.8, 0.6, 0.2}
\definecolor{red}{rgb}{0.8, 0.2, 0.2}
\definecolor{darkred}{rgb}{0.6, 0.1, 0.05}
\definecolor{blueish}{rgb}{0.0, 0.3, .6}
\definecolor{light_gray}{rgb}{0.7, 0.7, .7}
\definecolor{pink}{rgb}{1, 0, 1}
\definecolor{greyblue}{rgb}{0.25, 0.25, 1}

\usepackage{blindtext}

\renewcommand{\paragraph}[1]{\vspace{1em}\noindent\textbf{#1}.}

\def\checkmark{\tikz\fill[scale=0.4](0,.35) -- (.25,0) -- (1,.7) -- (.25,.15) -- cycle;}
\newcommand{\xmark}{\ding{55}}%
\usepackage{ulem}\usepackage{pifont}

\usepackage{tikz}

\newcommand{\Daux}{\mathbf{D}^{aux}}
\newcommand{\Dpraux}{\mathbf{\tilde{D}}^{aux}}
\newcommand{\DprA}{\mathbf{\tilde{D}}^A}
\newcommand{\DA}{\mathbf{D}^A}
\newcommand{\X}{\mathbf{X}}
\newcommand{\Xpr}{\mathbf{\tilde{X}}}
\newcommand{\Y}{\mathbf{Y}}
\newcommand{\Ypr}{\mathbf{\tilde{Y}}}

\newcommand{\YS}{\mathbf{Y}_{S}}
\newcommand{\YSpr}{\mathbf{\tilde{Y}}_S}
\newcommand{\ZNS}{\mathbf{Z}_{NS}}

\newcommand{\z}{\mathbf{z}}

\newcommand{\zs}{\mathbf{z}_S}
\newcommand{\zspr}{\mathbf{\tilde{z}}_S}
\newcommand{\zns}{\mathbf{z}_{NS}}
\newcommand{\ZS}{\mathbf{Z}_{S}}
\newcommand{\ZSpr}{\mathbf{\tilde{Z}}_S}
\newcommand{\x}{\mathbf{x}}
\newcommand{\xpr}{\mathbf{\tilde{x}}}
\newcommand{\y}{\mathbf{y}}
\newcommand{\ypr}{\mathbf{\tilde{y}}}
\newcommand{\ys}{\mathbf{y}_S}
\newcommand{\yns}{\mathbf{y}_{NS}}
\newcommand{\yspr}{\mathbf{\tilde{y}}_S}
\usepackage{arxiv}
\usepackage{booktabs}
\usepackage{amsfonts}
\usepackage{wrapfig}

\usepackage{floatrow}
\newfloatcommand{capbtabbox}{table}[][\FBwidth]

\author{ Abhishek Singh\thanks{MIT Media Lab}
	\And
	Ethan Garza\footnotemark[1]
	\And
	Ayush Chopra\footnotemark[1]
	\And
	Praneeth Vepakomma\footnotemark[1]
	\AND
	Vivek Sharma$^{\ast}$
	\And
	Ramesh Raskar\footnotemark[1]
}

\begin{document}

\maketitle
\begin{abstract}
We propose \textit{sanitizer}, a framework for secure and task-agnostic data release. While releasing datasets continues to make a big impact in various applications of computer vision, its impact is mostly realized when data sharing is not inhibited by privacy concerns. We alleviate these concerns by sanitizing datasets in a two-stage process. First, we introduce a \textit{global decoupling} stage for decomposing raw data into sensitive and non-sensitive latent representations. Secondly, we design a \textit{local sampling} stage to synthetically generate sensitive information with differential privacy and merge it with non-sensitive latent features to create a useful representation while preserving the privacy. This newly formed latent information is a task-agnostic representation of the original dataset with anonymized sensitive information. While most algorithms sanitize data in a task-dependent manner, a few task-agnostic sanitization techniques sanitize data by censoring sensitive information. In this work, we show that a better privacy-utility trade-off is achieved if sensitive information can be synthesized privately. We validate the effectiveness of the sanitizer by outperforming state-of-the-art baselines on the existing benchmark tasks and demonstrating tasks that are not possible using existing techniques.
\end{abstract}


\section{Introduction}
\label{sec:intro}
Releasing datasets has resulted in methodological advancements in computer vision~\cite{deng2009imagenet,su2021affective} and machine learning (ML). However, the advancement is still limited by datasets that comply with modern privacy standards. The goal of this work is to alleviate privacy concerns in dataset release when it contains sensitive information. Since datasets are released independently of downstream tasks, we focus on the problem of protecting sensitive information in a task-agnostic manner. We refer to this problem as \textbf{sanitization}. Releasing sanitized datasets could galvanize the research community to make progress in the areas where raw data access is not feasible. 

As a motivating example, consider a hospital with a dataset of face images where ``ethnicity" and ``age" of every face is a sensitive detail. The hospital is enabled to share the dataset with untrusted parties for several applications if we can \textit{sanitize} all images in the dataset. To understand the benefit of sharing the dataset, we list the following use-cases that also motivate our experiments in Section~\ref{sec:exp}.\\
\textbf{UC1:} A crowd-sourcing company can build a facial recognition model for medical diagnostics~\cite{loos2003computer,stephen2017facial,chen2018development} from the sanitized dataset. This model will be deployed on cloud, therefore the prediction will be performed over sanitized images.\\
\textbf{UC2:} A group of researchers can develop a model of capturing keypoints from face images. Unlike UC1, they want the model to predict over unsanitized images. Hence sanitized images should be \underline{photo-realistic} to prevent a domain mismatch.\\
\textbf{UC3:} The hospital wants to share a sanitized dataset with a company to build an ML model to predict ``age". Similar to UC2, the hospital would perform prediction on unsanitized images hence the sanitized dataset should be photo-realistic. However, unlike UC2, prediction attribute ``age" is also a sensitive attribute requiring privacy.

Since there can be many such use-cases, it is impractical to assume that the hospital knows all use-cases in advance before releasing the dataset. Therefore, the goal of \textit{Sanitizer} is to transform the dataset by anonymizing sensitive information without the knowledge of the downstream use-cases. In addition to learning ML models, being task agnostic allows \textit{sanitizer} to do inference queries on sensitive datasets such as counting the number of faces with ``smiling" attribute, or counting X-ray images with ``lung cancer". Trivially cropping the sensitive parts from the image is not feasible because the pixels that reveal the sensitive information are present everywhere in a face image. Furthermore, unlike face images, identifying sensitive information visually may not be possible.
For instance, several recent works~\cite{korot2021predicting,kumar2021cardiac,betzler2021gender,banerjee2021reading,yi2021radiology} show ways in which sensitive demographics can be leaked leak from biomedical images using ML models.
\begin{figure*}[t]
 \centering
 \includegraphics[width=0.9\linewidth]{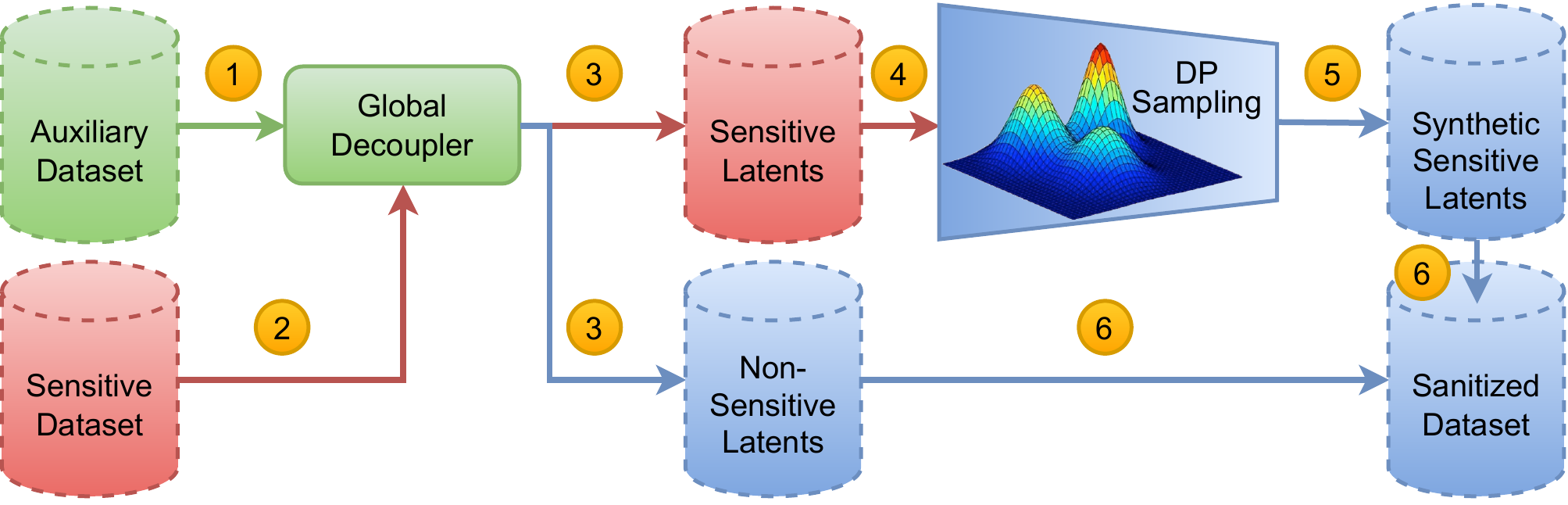}
 \caption{\textbf{Sanitizer pipeline} First, we learn a latent model (global decoupler) of the data distribution using non-sensitive auxiliary dataset (in \textcolor{green}{green}). Next, we use the latent model to decouple sensitive (in \textcolor{red}{red}) and non-sensitive (in \textcolor{blue}{blue}) information from the sensitive dataset to learn the distribution of sensitive latents. We synthetically generate sensitive latents by sampling from the distribution. Finally, we get the sanitized dataset by combining non-sensitive and synthetically generated sensitive latents.}
 \label{fig:pipeline}
\end{figure*}

Many works in sanitizing data have a different scope from the one considered here. This is either due to the notion of i) sensitive information or ii) utility. Typically identity~\cite{dwork2014algorithmic} of individuals is treated as sensitive information. While this notion protects privacy, we \textit{only} focus on a specific set of sensitive attributes. For example - in UC1, it is acceptable to share face images as long as ``ethnicity" and ``age" can be protected.  In works that do consider specific sensitive attributes, their notion of utility is typically task-dependent as in ~\cite{osia2018deep,osia2020hybrid,li2019deepobfuscator}. Although \textit{sanitizer} can be used for such problems due to its task-agnostic approach, not exploiting the knowledge about downstream tasks comes at the expense of a relatively lower utility. Existing works specifically in sanitization~\cite{gap,tiprdc} protect sensitive information by \textit{censoring} it. Unlike censoring based approaches, our main idea is to share synthetically generated sensitive information. While the data receiver can not infer the original sensitive information, our approach allows them to learn from anonymized sensitive information.

To design \textit{sanitizer}, we posit that sensitive data can be anonymized by replacing sensitive information with a synthetic one. However, for images, this synthetic replacement is not trivial to perform since the sensitive information is not localized in a region and sensitive attributes and non-sensitive attributes can share the same parts of data (ex. - race and gender). Therefore, we introduce a \textit{latent model} that exclusively isolates sensitive information into a smaller subspace. We learn the latent model using publicly available datasets and then use the model to isolate the sensitive information from the sensitive dataset. Next, we learn a generative model of the isolated sensitive information and synthesize sensitive latents by sampling from the model. We merge these samples with the non-sensitive latent representation to obtain sanitized data. We visualize the whole pipeline in Figure~\ref{fig:pipeline}.

\textbf{Contributions}: \textit{First,} we introduce a joint optimization framework for isolating sensitive information from data. \textit{Second}, we design a mechanism for anonymizing sensitive image datasets. \textit{Third}, we empirically demonstrate various applications of sanitization to show the benefit of \textit{sanitizer} over existing approaches.
\section{Problem Formulation}
\label{sec:formulation}
\textbf{Terminology}: Consider a data holder $A$ with access to a dataset $\DA=\{\X,\Y\}$ with $N$ data points. Let $\x \in \X$ and $\y \in \Y$ represents a pair of sample and \textit{set} of labels ($\x,\y$) describing distinct attributes of $\x$. For instance, if $\x$ is a face image of an individual, the \textit{set} $\y$ may include the age, gender, and ethnicity of the individual. For $A$, certain attributes in the label set $\y$ represent sensitive information (called $\ys$) while others are non-sensitive ($\yns$) such that $\y = \{ \ys \cup \yns \}$. These sensitive attributes are $A$'s secrets that prevent $A$ from sharing $\DA$. While $A$ can release $(\x,\yns)$, an attacker can guess $\ys$ using $\x$ by exploiting correlation between $\x$ and $\ys$. Hence, to release $\DA$ for arbitrary downstream tasks, \underline{sanitization techniques} transform every sample in $\DA$ from $(\x,\y)$ to $(\xpr, \ypr)$ resulting in sanitized dataset $\DprA$ that can be shared with untrusted parties. The key challenge is anonymizing sensitive information while maximally retaining the utility. We assume that an auxiliary dataset $\Daux$ from the same distribution as $\DA$ is publicly accessible to all parties.

\textbf{Sanitizer Overview}: We perform sanitization in a two stage process: i) \textit{Global decoupling} and ii) \textit{Local sampling}. \textit{Global decoupling} stage learns a latent model (parameterized by $\theta, \phi$) of data using $\Daux$ for decoupling raw data ($\x$) into sensitive ($\zs$) and non-sensitive ($\zns$) latents. We assume that the auxiliary dataset and the sensitive dataset come from the same distribution $p(\x,\y)$. This stage does not require access to $A$'s dataset ($\DA$) and hence can be performed independently making this stage \textit{global} since the same model can be utilized by different sensitive-data owners. We discuss the design of the \textit{global decoupler} in Section~\ref{sec:global_decoupler}. \textit{Local sampling} stage learns a generative model ($f(\psi,\cdot)$) of sensitive latents in $\DA$. We obtain the sensitive latents in $\DA$ using the \textit{global decoupler} from the first stage. Finally, we obtain the sanitized dataset by merging every non-sensitive latent with independently sampled sensitive-latent. We discuss the \textit{local sampling} stage in Section~\ref{sec:mechanisms}.

\textbf{Threat Model:} We assume that the \textit{untrusted} data-receiver can act as an attacker by utilizing auxiliary dataset $\Daux$, parameters of \textit{global decoupler} ($\theta, \phi$) and \textit{local sampler} ($f(\psi,\cdot)$) as a side information. The side information allows the attacker to generate a mapping between $\Daux$ and its sanitized version $\Dpraux$. The attacker's goal is to recover $A$'s sensitive attribute $\ys$ from the sanitized dataset $\DprA$. Since $\Daux$ and $\DA$ come from the same distribution, the attacker can model the problem of inferring the sensitive attributes as an ML problem. By learning a mapping between sanitized samples ($\xpr\in\Xpr^{aux}$, $\ypr\in\Ypr^{aux}$) and sensitive attributes ($\ys\in\YS^{aux}$) using $\Daux$, the attacker can attempt inferring sensitive attributes from $\DprA$. This threat model is different from differential privacy~\cite{dwork2014algorithmic} which seeks to protect identifiability.

\textbf{Defining information leakage:} Information leakage for sanitization has been typically defined statistically~\cite{dwork_calibrating_2006,makhdoumi2013privacy,sankar2010information}; however, estimating these statistics requires estimation of probability distributions making it intractable for non-linear queries over higher-dimensional datasets (images). Alternatively, leakage can be quantified by simulating an attacker's performance by making some assumptions. The goal of sanitization is to minimize the distinguishability of the original sensitive attributes $\ys$ from other possible values (domain($\ys$)) sensitive attributes can take. For example, if ``ethnicity" is a sensitive attribute then, informally, leakage is the likelihood of the attacker's correct \textit{estimate} about the race of the sanitized face image. Formally, this can be modeled by a change in belief over the sensitive attribute before (prior $p(\ys)$) and after (posterior $p(\ys|\xpr)$) observing the sanitized sample ($\xpr,\ypr$). A similar notion is formalized in the pufferfish framework~\cite{kifer2014pufferfish} where the quantity $\frac{p(\y_{s_j}|\xpr, \theta)}{p(y_{s_i}|\xpr, \theta)}/\frac{p(y_{s_j})}{p(y_{s_i})}$ for all possible sets of secrets ($y_{s_i}, y_{s_j}$) and for all possible priors on data $\theta$ is bounded by $e^\epsilon$ where $\epsilon$ is a privacy parameter. Note that satisfying this definition requires modeling various possible data evolution and attacker scenarios. We focus only on a single type of attacker described in the threat model and therefore use a data-driven approach to quantify leakage. This data-driven adversary learns the joint distribution $p(\Xpr, \YS)$ using the side information. Finally, the leakage of the sanitized datasets is evaluated as the difference between the accuracy of the adversary to correctly estimate the sensitive information $p(\ys|\xpr, \theta)$ and the estimation of an uninformed adversary $p(\ys|\theta)$. We note that existing works in sanitization~\cite{tiprdc,gap} use the same criterion to evaluate information leakage.

\textbf{Desiderata}: In both stages of \textit{sanitizer}, we have two desirable properties corresponding to privacy and utility; in total, we get four desirable properties that we elaborate on now. The first stage is \textit{global decoupling} where we learn to separate a sample $\x$ into sensitive and non-sensitive latent $\zs$ and $\zns$ respectively. Therefore, the desirable property \textbf{P1} requires $(\zs,\zns)$ to be independent. In other words, \textbf{P1} requires non-sensitive latent $\zns$ does not leak information about $\zs$. This property can be achieved trivially by sharing all zeroes therefore to enforce utility, we desire property \textbf{P2} that requires $p(\x|\zs,\zns)$ to be maximum. Property \textbf{P2} requires $(\zs,\zns)$ to be useful enough for describing the original sample $\x$. This completes the privacy-utility desiderata for the \textit{global decoupling} stage. 
The \textit{local sampling} stage focuses only on transforming the sensitive latent, therefore, both privacy and utility desideratum only focus on $\zns$. For the privacy desiderata \textbf{P3} in this stage, we require that sanitized sensitive latents $\zspr$ and $\zspr'$ obtained from $\x,\x'\in\X$ respectively are indistinguishable from each other. \textbf{P3} enforces that identifying original data sample based only on the sensitive information should not be possible, i.e. $p(\zspr\sim f_\psi(\zs)) = p(\zspr\sim f_\psi(\zs'))$. For the example of a face image with sensitive "ethnicity", \textbf{P3} requires synthetically generated $\zspr$ should be independent of the original "ethnicity" of the sample. We can trivially solve \textbf{P3} by sharing only zeroes, therefore to ensure utility, we introduce property \textbf{P4} that requires the \textit{distribution} of original sensitive and synthetic sensitive latents to be the same. Specifically, the property implies $p(\zs)=p(\zspr)$. Next, we model these desiderata to design our technique.

\section{Method}
\label{sec:method-vae}
\label{sec:method}

\begin{wrapfigure}{r}{0.6\columnwidth}
\centering
\includegraphics[width=0.99\linewidth]{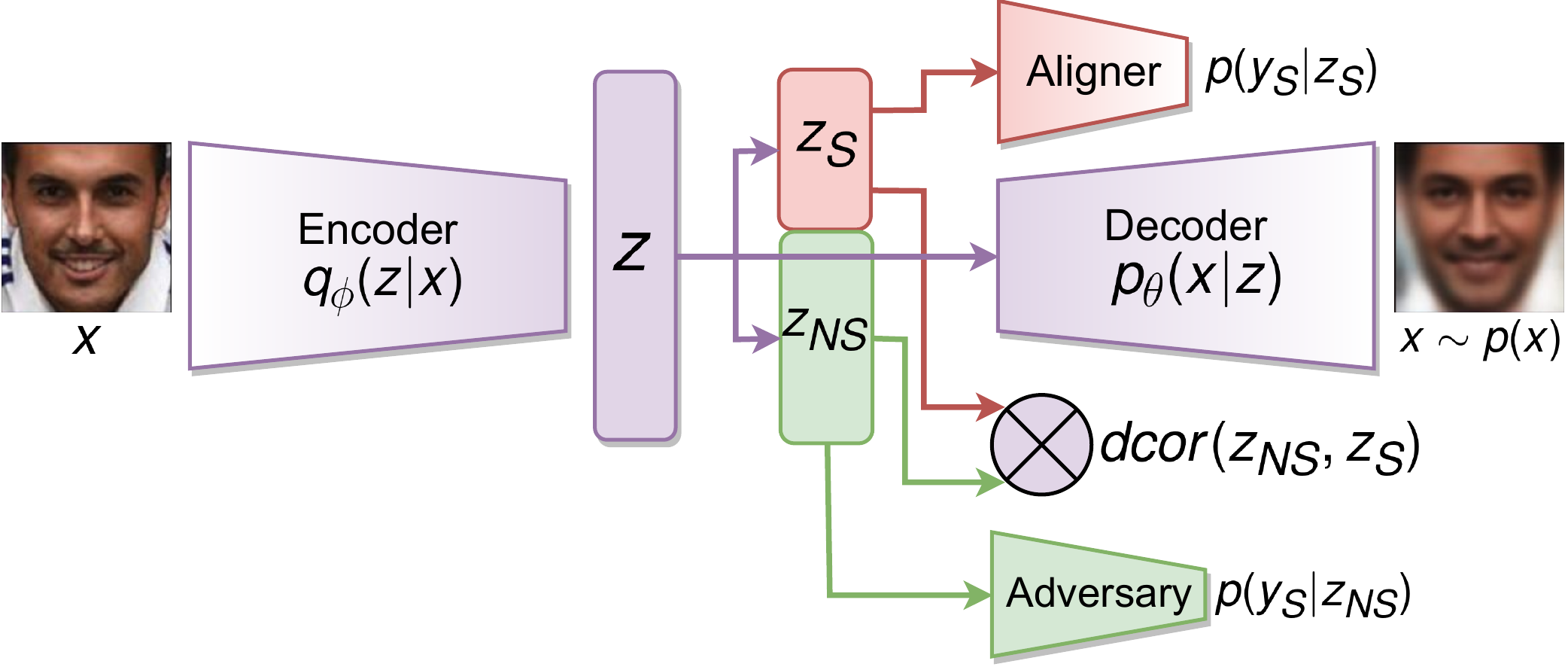}
     \caption{Architecture for the proposed global-decoupler. The encoder is used to sample $\z~\sim q_\phi(\z|\x))$ which is partitioned into $(\zs,\zns)$. We use aligner to encourage $\zs$ to carry information relevant to $\ys$. We use adversary to encourage $\zns$ does not carry information about $\ys$ and we utilize distance correlation to minimize correlation between $\zs$ and $\zns$ to prevent $\zs$ from carrying information about $\yns$. Finally, we train the encoder and decoder blocks to maximize the likelihood as done in VAE~\cite{vae}.\vspace{-0.7cm}}
\label{fig:arch}
\end{wrapfigure}

We sanitize a sensitive dataset in a two-stage process. 
The first stage is \textit{global decoupling} where we learn a latent model of the data distribution using auxiliary dataset $\Daux$. 
Our goal is to learn a latent model of data that maximally \textit{decorrelates} $\zs$ and $\zns$ for every sample $\x$ (P1) and \textit{preserves} all details of the sample in $\z$ (P2).
We achieve this goal by designing \textit{global-decoupler} in Section~\ref{sec:global_decoupler}.
The second stage is \textit{local sampling} where we learn the distribution of the sensitive portion $\ZS$ of our sensitive dataset $\DA$. 
Our goal is to sample from the distribution $\ZSpr$ such that estimating original sensitive attribute $\YS$ is not feasible (P3) and the distribution of $p(\ZS)$ and $p(\ZSpr)$ is similar (P4). We achieve this goal by designing \textit{DP-sampling} mechanism in Section~\ref{sec:mechanisms}.
We summarize the overall pipeline in Figure~\ref{fig:pipeline}. 
\subsection{Global decoupling for isolating sensitive information}
\label{sec:global_decoupler}
Our goal is to design a latent model of the data distribution $p(\x,\y)$ such that $\x$ can be decoupled into latents $\zs$ and $\zns$. We call this latent model as \textit{global decoupler}. We design it by integrating four components - i) generative model, ii) aligner, iii) decorrelator and iv) adversarial training. We describe the architecture of \textit{global-decoupler} in Figure~\ref{fig:arch}. Generative models (~\cite{goodfellow2014generative,vae,rezende2015variational}) are being increasingly used to perform such latent modeling . Specifically, we build upon VAE ~\cite{vae} since they provide the flexibility of modeling data with constraints on the probability density of the latent space $p(\z)$. Given a dataset $\X$, VAEs ~\cite{vae,rezende2014stochastic} model the distribution of samples $p(\x), \forall \x\in \X$ by learning parameters $\phi$ of approximate posterior $q_\phi(\z|\x)$ and $\theta$ for the likelihood $p_\theta(\x|\z)$. $\beta$-VAE \cite{betavae} improves the disentanglement between the components of $\z$ sampled from $q_\phi(\z|\x)$ by regularizing the KL divergence between the prior $p(\z)$ and approximate posterior $q_\phi(\z|\x)$. To improve disentanglement between every $\z_i$, existing works such as Factor-VAE\cite{factorvae} and TCVAE\cite{tcvae} regularize the total correlation of $q(\z)$ measured by $\mathsf{KL}(q(\z)||\prod_{i=1}^{m}q(\z_i))$ where $\mathsf{KL}$ refers to the KL divergence and $m$ is the total number of components of $\z$. However, high degree of disentanglement between \textit{every component} can hinder the reconstruction quality~\cite{betavae}. Therefore, instead of disentangling every pair of $(\z_i,\z_j)$, we propose a new regularized \textit{global-decoupler} to focus on the disentanglement of $(\zs,\zns)$ instead.

A key characteristic of VAE is that the decoupled latent representations are unordered. Hence, there is no explicit control on which dimensions encode what semantic attributes. This is a challenge for our work that ideally requires that representations encoding the sensitive attributes be contiguous for decoupling. Intuitively, we decouple the vector $\z\sim q_\phi(\z|\x)$ into $\zs$ and $\zns$ with additional regularization constraints that encourage independence between $\zs$ and $\zns$. Formally, we reformulate the original VAE objective with an \textit{aligner} $g_u(\cdot)$ parameterized by $u$ to estimate $\ys$ from $\zs$, the intuition is that the \textit{aligner}'s gradient flow will encourage $q_\phi(\cdot)$ to maximize relevant information between $\ys$ and $\zs$. Since all latents are known to be correlated with each other to a certain extent, we need to prevent leakage of $\ys$ in $\zns\sim q(\zns|\x)$. Unlike FactorVAE~\cite{factorvae} or TCVAE~\cite{tcvae} that regularize the total correlation disentangling each dimension, we propose to regularize correlation between sensitive($q(\zs)$) and non-sesntive($q(\zns)$) latents. We re-formulate the objective for $q_\phi(\cdot)$ to minimize distance correlation~\cite{szekely2007measuring} between $q(\zs)$ and $q(\zns)$. To motivate the use of distance correlation, we note that directly estimating probability density is intractable for high dimensional representations, various measures such as HSIC~\cite{gretton2005measuring}, MMD~\cite{borgwardt2006integrating} and distance correlation~\cite{vepakomma2020nopeek} are used. Distance correlation between $n$ samples of two vectors $\x$ and $\y$ can be obtained as following:
$$dcorr(\x,\y) = \frac{dcov(\x, \y)}{\sqrt{dcov(\x, \x)*dcov(\y, \y)}}$$ where $dcov()$ is the sample distance covariance analogue of covariance defined as $dcov(\x,\y) = \frac{1}{n^2}\sum_{j=1}^{n}\sum_{k=1}^{n}\hat{\x}_{j,k}\hat{\y}_{j,k}$. Here $\hat{\x}$ and $\hat{\y}$ are obtained by computing double centered euclidean distance matrices of $\x$ and $\y$. In particular, we use distance correlation ($dcorr$) because it can measure nonlinear correlations between samples from random variables of arbitrary dimensions ($\zs$ and $\zns$ can have different dimensionality), allows for efficient gradient computation and does not require any kernel selection or parameter tuning, unlike HSIC and MMD. We do note that $dcorr$ is measured as a sample statistic and hence larger sample size is desirable for the unbiased sample statistic to represent the population notion of the distance correlation. To prevent information leakage of $\ys$ from $\zs$, we use a proxy attacker network $h_v(\cdot)$ that is trained adversarially to learn parameters $v$ which constrains $\zns$ to not carry information relevant to $\ys$. The final objective can be summarized as: \\
\begin{equation}
    L_1(\theta,\phi,\beta) = \mathbb{E}_{q_\phi (\z|\x)}[log p_\theta (\x|\z)] - \beta D_{KL}(q_{\phi}(\z|\x)||p(\z))
    \label{eq:betavae}
\end{equation}
\begin{equation}
    L_2(\phi, u)=\ell_1(g_u(\z_i\sim q_\phi(\x)|_{i\leq k}), \ys) 
    \label{eq:pred}   
\end{equation}
\begin{equation}
    L_3(\phi) = dcorr(\z_i\sim q_\phi(\x)|_{i\leq k}, \z_i\sim q_\phi(\x)|_{k<i\leq m})
    \label{eq:dcorr}
\end{equation}
\begin{equation}
 L_4(\phi, v) = \ell_2(h_v(\z_i\sim q_\phi(\x)|_{k<i\leq m}), \ys)
 \label{eq:adv}
\end{equation}
Here $k$ and $m$ are the dimensionalities of vectors $\zs$ and $\z$, respectively. $L_1$ is the $\beta$-VAE~\cite{betavae} formulation of VAE's evidence lower bound where the parameter $\beta$ encourages disentanglement between every component of $\z$. Increasing $\beta$ favors the property P1 (by encouraging independence) but hurts the property P2 (by reducing reconstruction). $L_2$ is the objective for training the parameters of the aligner model. However, $L_1$ does not prevent $\zns$ from leaking information about $\ys$. Hence, we optimize $L_4$ adversarially to prevent information leakage. Finally, we minimize distance correlation between $\zs$ and $\zns$ to prevent $\yns$ from encoding information about $\zs$ and encourage decoupling $\zs$ and $\zns$. \underline{Jointly optimizing $L_1$, $L_2$, $L_3$ and $L_4$ helps achieve properties P1 and P2}. We validate each component's benefit via ablation studies in Section~\ref{sec:discussion}.

$\ell_1, \ell_2$ can be cross-entropy or $\ell_p$-norm (often $p=2$) depending upon $\ys$. The parameters $\phi,\theta,u,v$ are trained jointly with following objective:
\begin{equation}
 \min_{\theta, \phi, u} \alpha_1 L_1 (\theta,\phi,\beta) + \alpha_2 L_2 (\phi, u) + \alpha_3  L_3(\phi) - \alpha_4 \min_{v} L_4(\phi, v)
 \label{eq:joint}
\end{equation}
where $\beta,\alpha_1,\alpha_2,\alpha_3,\alpha_4$ are scalar hyper-parameters that yield a trade-off between the privacy (property P1) and utility (property P2). We reiterate that this stage only accesses auxiliary dataset $\Daux$ for the training and evaluation. Hence the parameters of the global-decoupler do not leak any sensitive information. 

\subsection{Local sampling for synthesizing sensitive latents}
\label{sec:mechanisms}
In this stage, we design the \textit{DP-sampling} mechanism to sanitize every sample in the sensitive dataset. A sanitized sample $(\xpr,\ypr)$ is obtained from $(x,y)\in \DA$ by extracting the sensitive and non-sensitive latents ($\zs,\zns$) from the global-decoupler and replacing the sensitive latent with a synthetic one ($\zspr$). To satisfy the privacy desiderata of our second stage P3, $\zspr$ is sampled independently of $\zs$. In contrast to prior works~\cite{huang_context-aware_2017,li2020tiprdc,deeprotect} that focus on censoring sensitive attributes, a key benefit of our mechanism is in anonymizing the sensitive information for individual data points while enabling downstream tasks that may benefit from joint distribution $p(\zs, \ys)$. Motivated by the use-case UC3, we demonstrate in Section~\ref{sec:exp} under experiment E5 on how to train an ML model on a sanitized dataset that predicts the sensitive attributes of unsanitized images. To motivate our DP-sampling mechanism, we first discuss a trivial suppression-based mechanism and a naive DP mechanism.

\textbf{a) Suppression}: The key is to explicitly remove sensitive information by replacing $\zs$ with a zero vector (i.e. $\zspr$ is a zero vector). While this approach censors the sensitive latent, it is not possible to learn the distribution $p(\x,\ys)$ under this mechanism resulting in a violation of our desired property P4. Therefore, we sanitize the sensitive information $\zs$ using DP mechanisms.

\begin{figure}
\centering
\includegraphics[width=0.59\columnwidth]{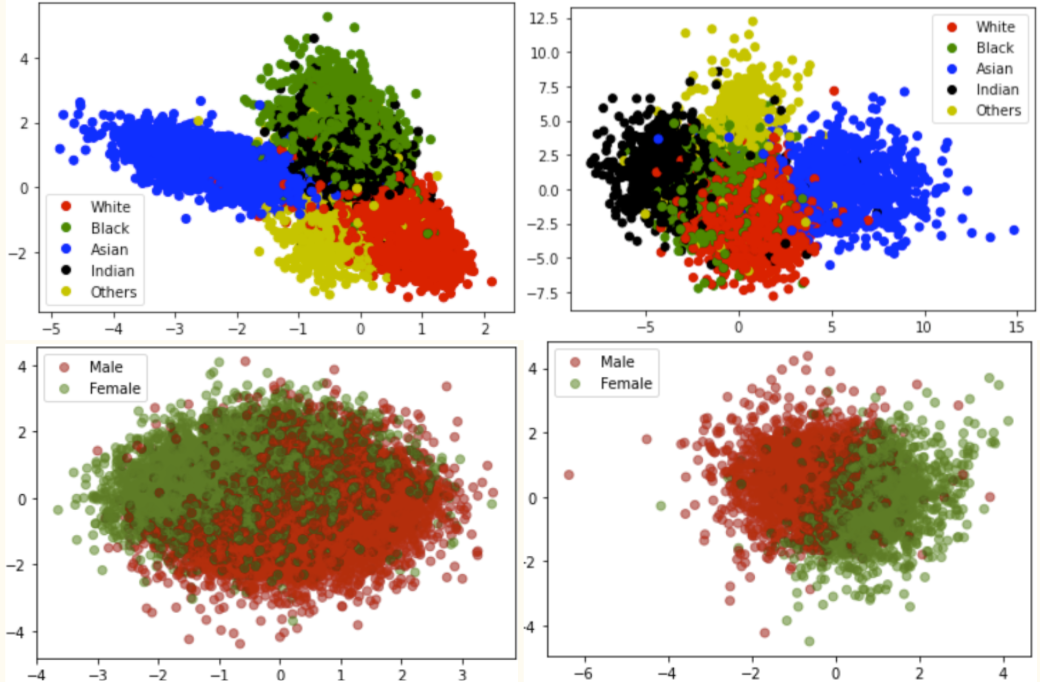}
\caption{\textbf{Latent space visualization of $\ZS$} by plotting its two components with the color of their sensitive attribute. We keep sensitive attributes as ``race" and ``gender" for the plots in the first and second row respectively using UTKFace~\cite{utkface}.}
    \label{fig:latentspace}
\end{figure}

\textbf{b) DP-Obfuscation}: The key idea is to add privacy calibrated noise to $\zs$. This calibration can be formalized using DP where a mechanism $\mathcal{M}$ is $\epsilon$-differentially private~\cite{dwork2014algorithmic} if for every neighboring datasets $X,X'$ and every output set $S\subset Range(\mathcal{M})$, the following inequality holds: $\mathbb{P}(\mathcal{M}(X)\in S) \leq e^\epsilon\mathbb{P}(\mathcal{M}(X') \in S)$. Here, we use the laplace mechanism~\cite{dwork_calibrating_2006} that adds noise sampled from a laplace distribution with variance as the $\ell_1$-sensitivity of the query $q$. In the context of our work, the $\ell_1$ sensitivity is defined as an identity function. Hence, to bound the sensitivity, we fix a pre-defined range $[a,b]\in\mathbb{R}$ in which $\zs$ can lie, giving us sensitivity as $||a - b||_1$. Any data sample $\x$ that results in $\zs\notin[a,b]$ is truncated to the closest vector in the range. To summarize, the mechanism can be described as $f(\zs)=\zs + \Delta$ where $\Delta$ is sampled from a laplace distribution, i.e. $\Delta\sim Lap(0, \frac{||a-b||_1}{\epsilon})$. While this approach allows us to have a trade-off between the privacy property P3 and utility property P4 by controlling $\epsilon$, we get a sub-optimal trade-off due to two reasons: i) truncation step discards the values outside the range $[a,b]$ ii) the noise is added independently to every component of $\zs$ therefore throwing away the structure present in the distribution of $\zs$ as shown in Figure~\ref{fig:latentspace}. This motivates developing a more utility conducive mechanism that can utilize structure in the latent space of $\zs$.

\textbf{c) DP-Sampling}: Our goal is to utilize the structure present in the distribution of sensitive latents $p(\zs)$ as shown in Figure~\ref{fig:latentspace}. We can observe the points to be clustered around their respective sensitive attribute. Therefore, \textit{instead of adding uniform and independent noise, we propose to learn the distribution of $\zs$ and sample from it}. Since the data is low dimensional, learning a Gaussian mixture model~\cite{mclachlan2019finite} suffices to model the distribution. However, sampling from the mixture model could leak sensitive information since our threat model considers the parameters of the sampling model are accessible to the attacker. Therefore, we learn the covariance matrix of $\ZS$ and perturb it in a differentially private manner before sampling from it. If the learned model of the data satisfies DP then due to the post-processing invariance property~\cite{dwork2014algorithmic} of DP, the samples obtained from this model would also satisfy DP. We learn the GMM model (parameterized by $\psi$) using the sensitive dataset $(\ZS, \YS)$. Such a sampling scheme also provides the benefit of sampling labeled pairs $(\ZSpr, \YSpr)$ which is required for performing supervised learning.
We utilize RONGauss mechanism~\cite{chanyaswad2019ron} to learn a differentially private covariance matrix of the sensitive latent dataset $\ZS$. For each $\zs\in \ZS$, the mechanism performs random orthonormal projection to a lower-dimensional. We learn the mean and covariance for each category in a differentially private manner. Low dimensional projection improves utility by reducing the perturbation required for the same amount of privacy. Finally, we obtain synthetic sample $\zspr$ by sampling from the Gaussian model and reprojecting it back to the original dimensionality of $\zs$. Formally, this can be written as $\zspr, \yspr \sim f_\psi(\zspr, \yspr)$, here $\psi$ is learned using original sensitive dataset $\ZS$.  We note that, unlike standard GMM, here we use only a single mode from the GMM for every unique class. This is a more accurate description of the data since every sample $\zs$ is uniquely associated with a single sensitive attribute $\ys$. While the RONGauss mechanism learns and samples the whole data space for providing a uniform privacy guarantee, we only sample sensitive latents $\zs$ instead of $\z$ since the goal is to protect sensitive attributes and not uniform privacy. We note that this DP-sampling does not give a uniform privacy guarantee on $\ZSpr$ since information about sensitive attributes can leak from $\ZNS$ too. We developed global-decoupler to address this specific issue. Our proposed sampling scheme can be extended to synthetic data release~\cite{tao2021benchmarking} by treating every component in $\z$ as sensitive (i.e. $k=m$) and presents interesting future work.

\textbf{Remark}: In this section, we presented our two-stage sanitization process. The main advantage of separating \textit{sanitizer} as a two-stage process is that developing \textit{global decoupling} is a one-time procedure and can be performed by a third party that distributes the trained model to different data owners $A$'s which can apply sanitizing mechanisms individually. This modular process is an efficient way to release sensitive datasets if there are multiple $A$'s involved. Furthermore, we believe that future works can improve either of the two stages independently.
\section{Experiments}
\label{sec:exp}
\begin{figure*}[t]
\centering
\begin{minipage}[t]{0.31\linewidth}
    \centering
     \includegraphics[width=1\columnwidth]{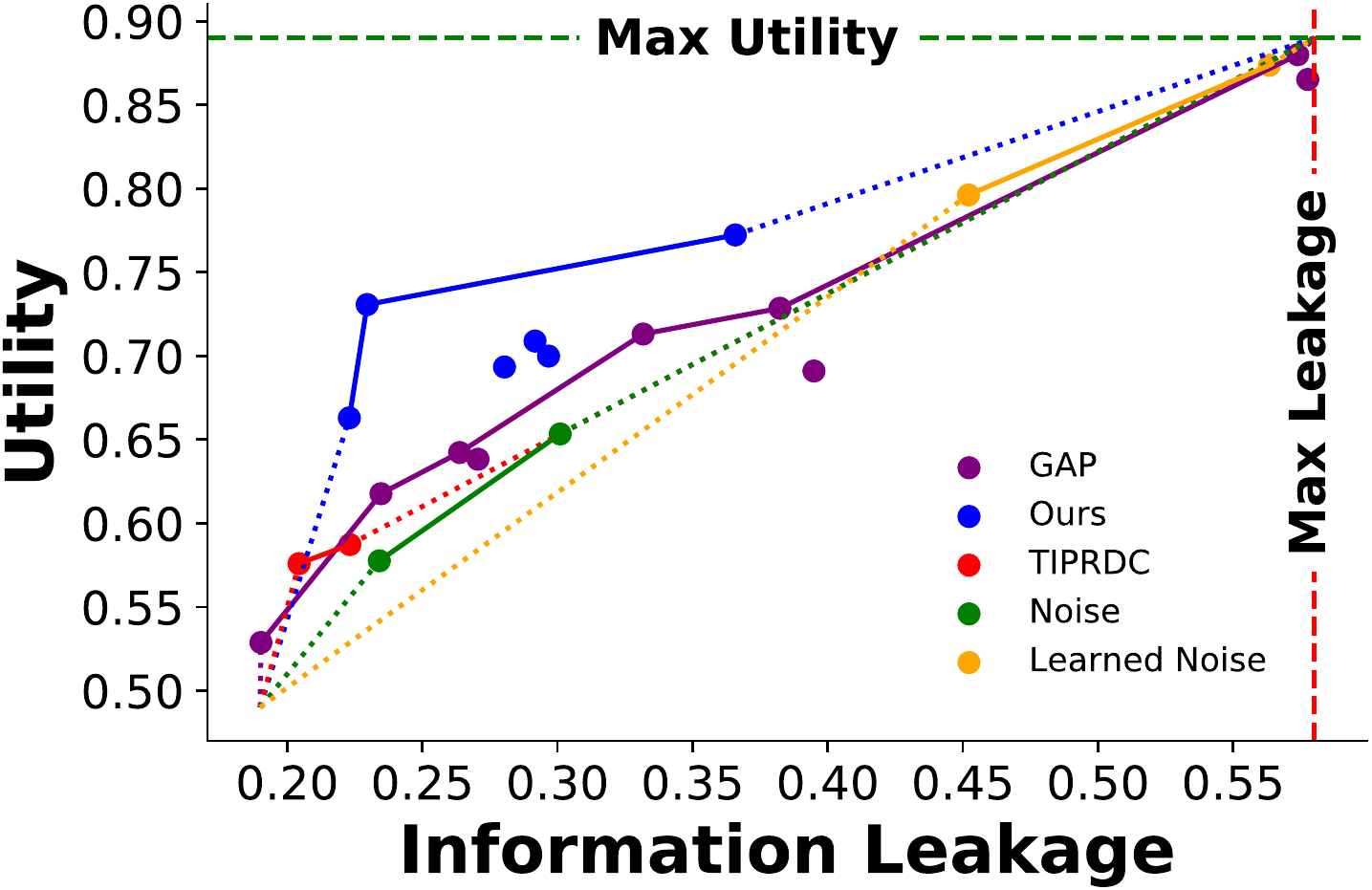}
     \centerline{ (a) FairFace-R-G }
     \label{fig:fairface}
\end{minipage}
\hspace{0.1cm}
\begin{minipage}[t]{0.31\linewidth}
    \centering
    \includegraphics[width=1\columnwidth]{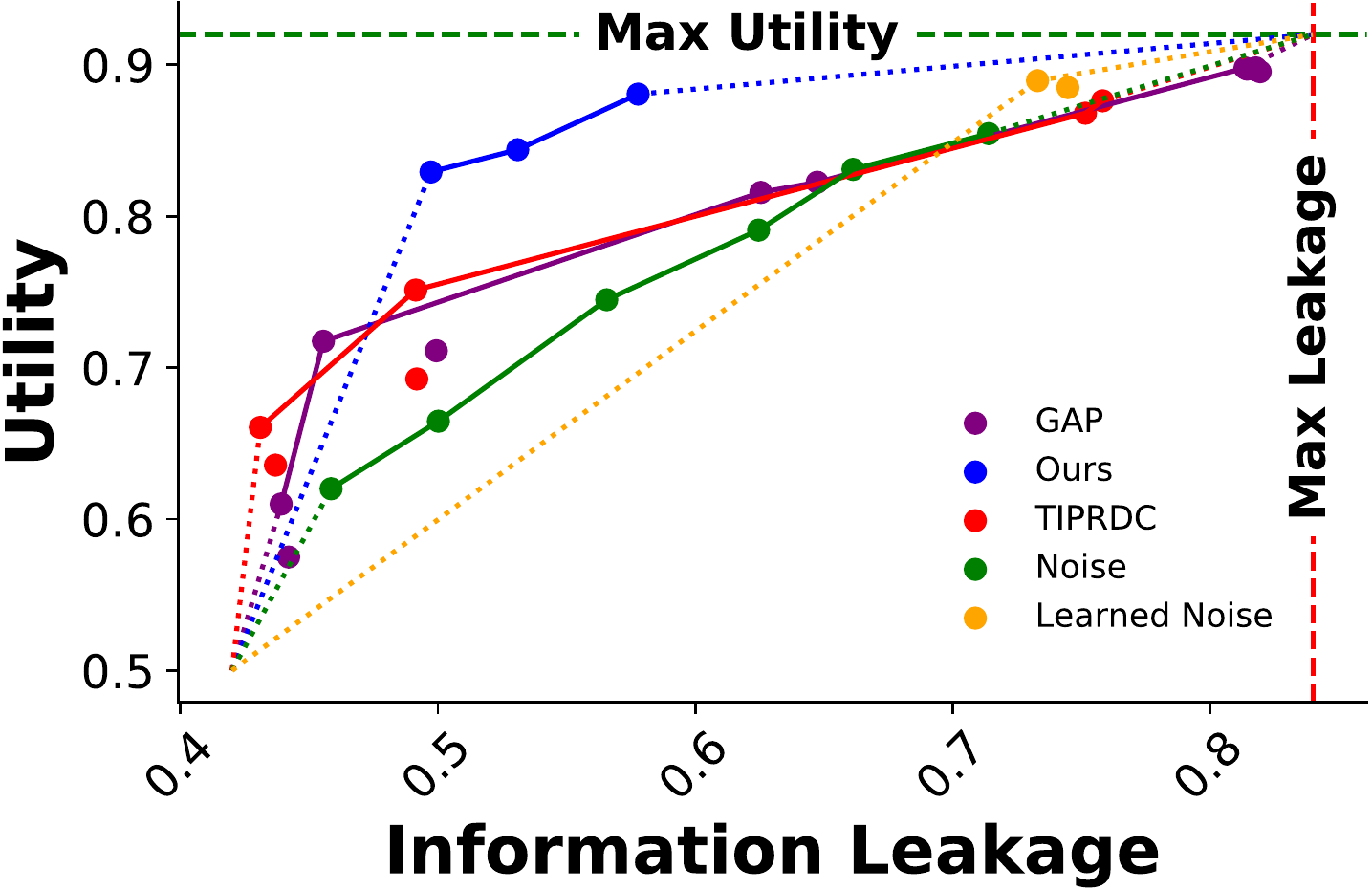}
    \centerline{ (b) UTKFace-R-G }
    \label{fig:utkface}
\end{minipage}
\hspace{0.1cm}
\begin{minipage}[t]{0.31\linewidth}
    \centering
    \includegraphics[width=\columnwidth]{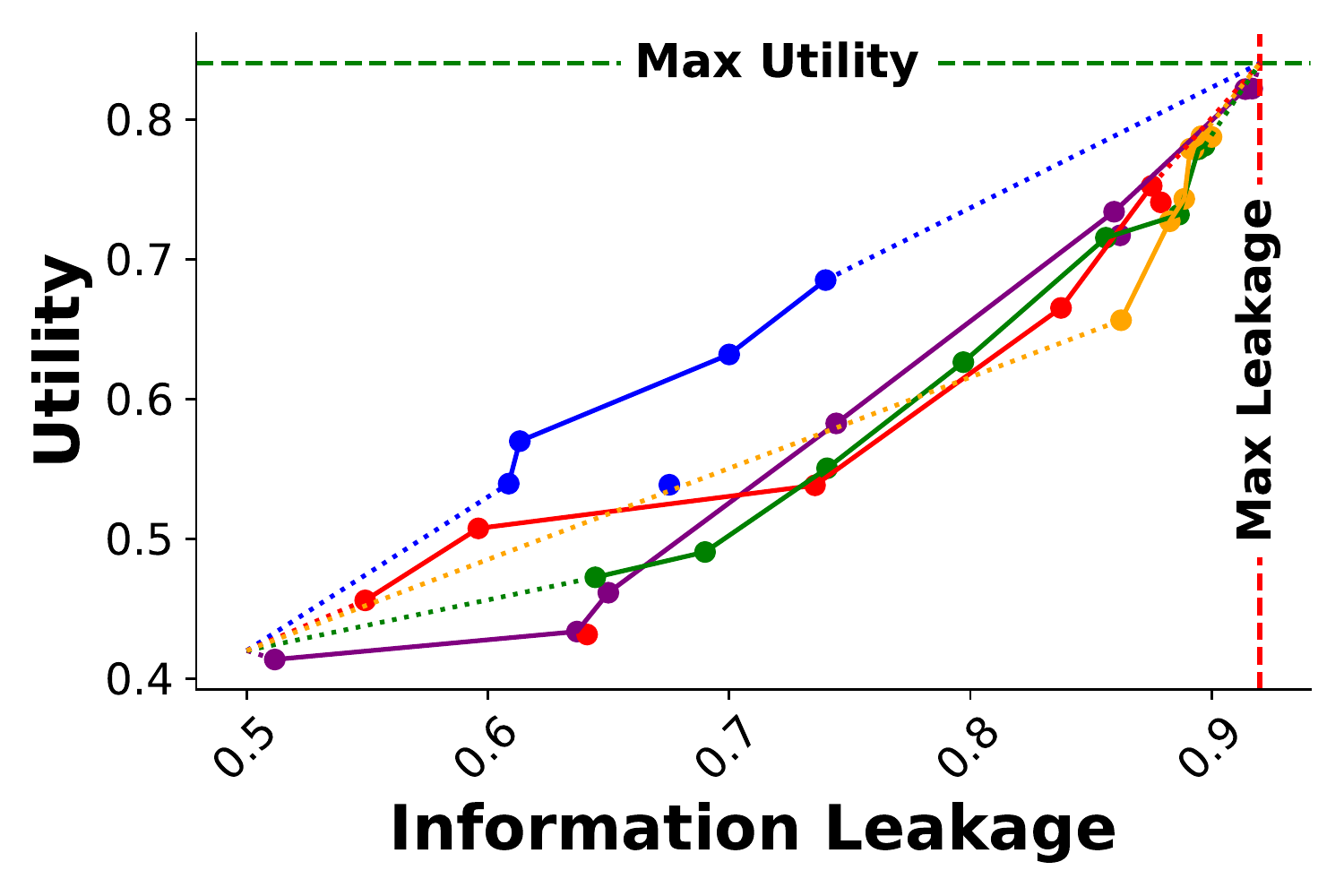}
    \centerline{ (c) UTKFace-G-R}
    \label{fig:utkface_g}
\end{minipage}

\begin{minipage}[t]{0.31\linewidth}
    \centering
    \includegraphics[width=1\columnwidth]{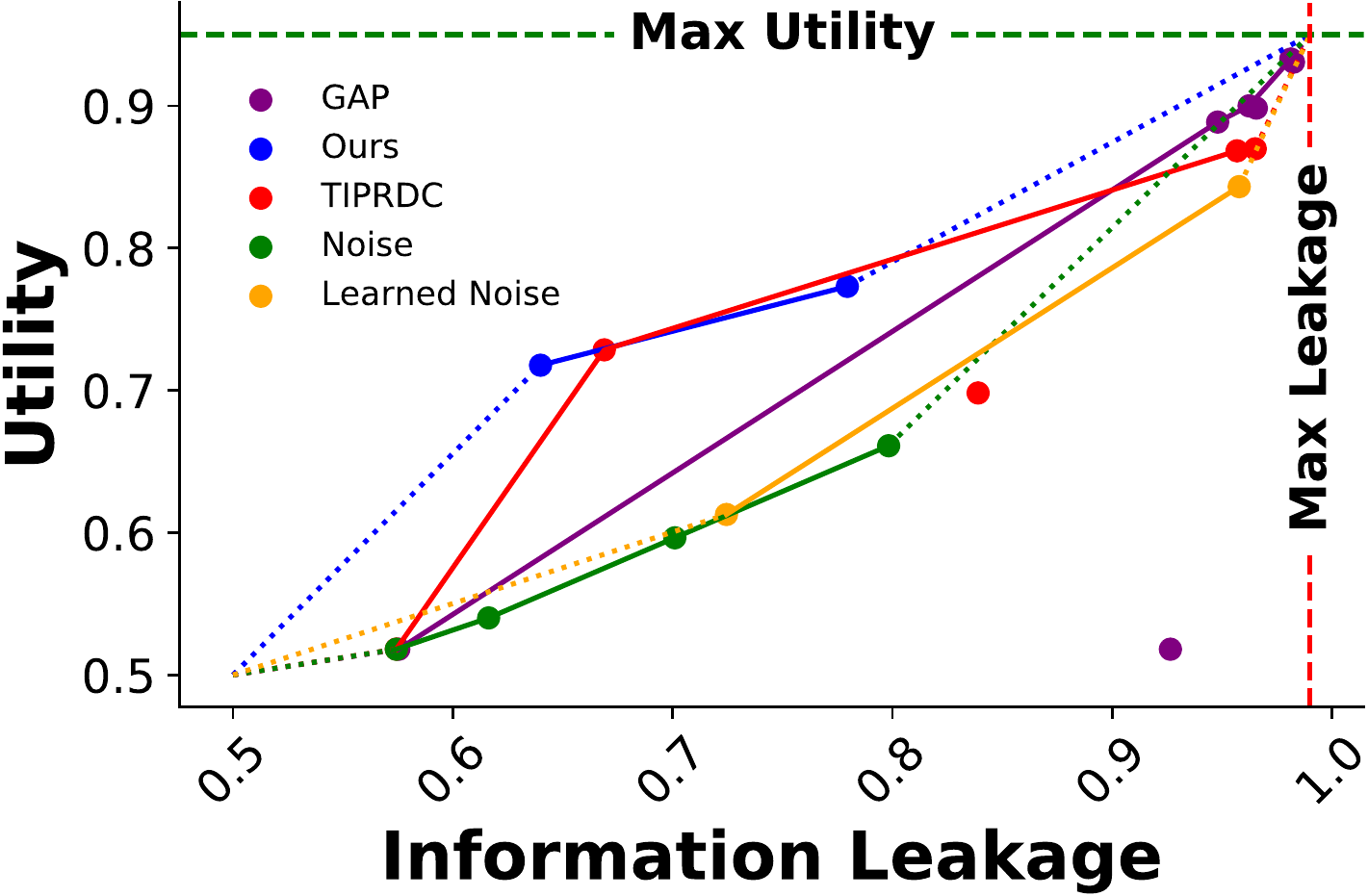}
    \centerline{ (d) CelebA-G-M }
    \label{fig:celeba}
\end{minipage}
\hspace{0.1cm}
\begin{minipage}[t]{0.31\linewidth}
    \centering
     \includegraphics[width=1\columnwidth]{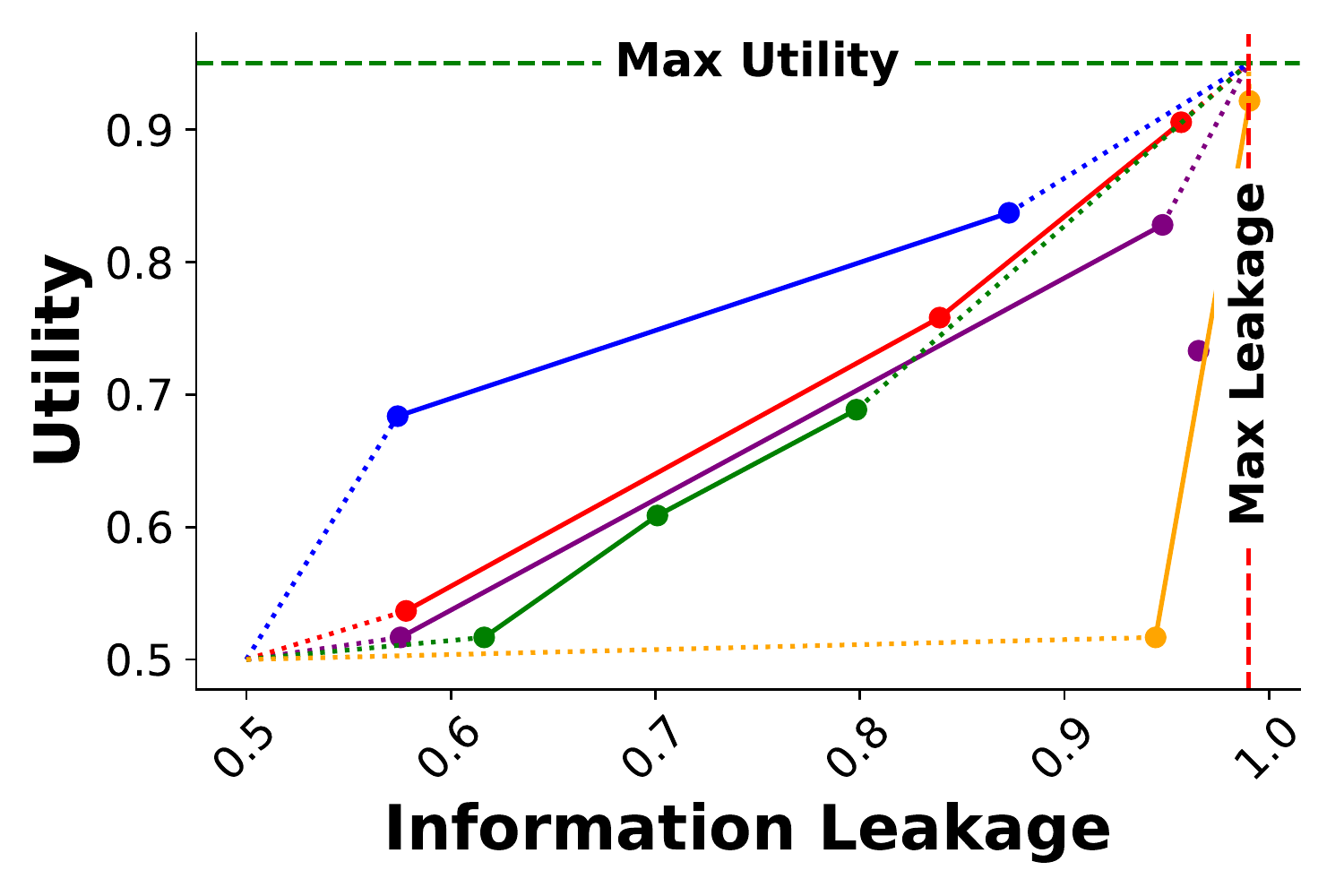}
     \centerline{ (e) CelebA-G-S }
     \label{fig:celeba-H}
\end{minipage}
\hspace{0.1cm}
\begin{minipage}[t]{0.31\linewidth}
    \centering
    \includegraphics[width=1\columnwidth]{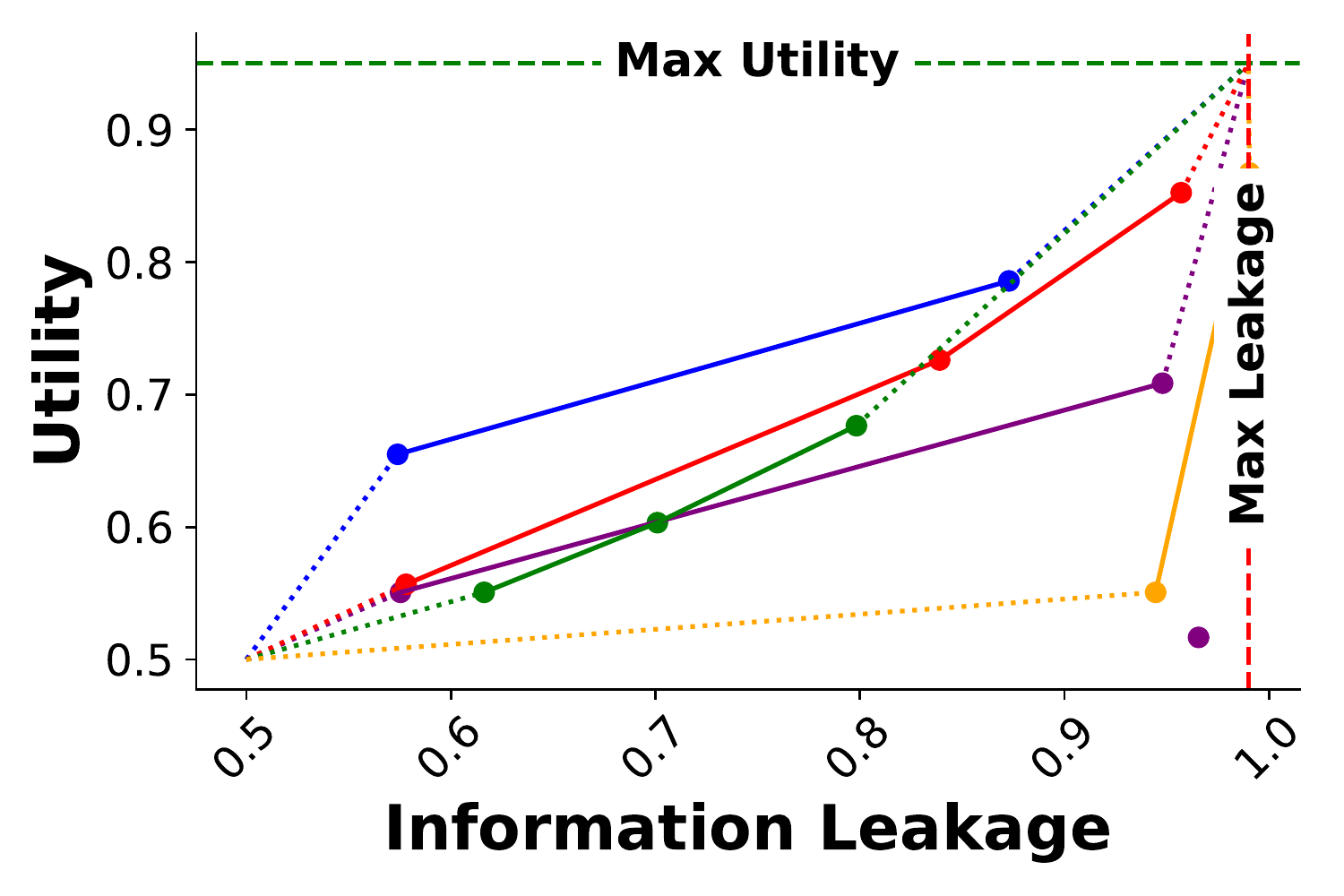}
    \centerline{ (f) CelebA-G-C }
    \label{fig:celeba-E}
\end{minipage}
\caption{\textbf{E1: Privacy-utility trade-off evaluation on different datasets:} We plot sensitive information leakage as a proxy for privacy and one of the task attributes as a measure of utility for the sanitized dataset. Each point in this plot corresponds to training a \textit{sanitizer} model and then evaluating its performance by training the adversary model and utility model on the sanitized dataset. \textit{Sanitizer} performs better than all existing methods on all three datasets. Solid line represents the pareto-optimal curve for different methods. The dotted lines are extrapolation towards lowest and highest utility and leakage that is achievable trivially.}
\label{fig:tradeoff}
\end{figure*}
In this Section, we compare \textit{sanitizer} with different baselines under multiple experimental setups. Each experimental setup is focused on simulating a unique use case. For all experiments, we use CelebA~\cite{celeba}, UTKFace~\cite{utkface} and Fairface~\cite{fairface}. We split the dataset into $\Daux$ for the first stage and $\DA$ for the second stage. First, we train all techniques using $\Daux$ and obtain sanitized dataset. Then we perform leakage assessment by training an adversary to learn a mapping between sanitized dataset and sensitive information. We discuss the rationale for such an adversary in Section~\ref{sec:formulation}. Finally, we evaluate the utility of the sanitized data based upon the experimental setup. Since our goal is task-agnostic data release, the \textit{utility attribute is only used} after the sanitized dataset is released.
\subsection{Baselines and Evaluation}

\textbf{Baselines:} We compare against state-of-the-art visual sanitization techniques GAP~\cite{huang_context-aware_2017} and TIPRDC~\cite{li2020tiprdc}, and introduce new baselines for exhaustive comparison. \textit{GAP~\cite{huang_context-aware_2017}:} is trained adversarially to maximize loss for a proxy adversary trying to infer sensitive attributes on the sanitized images. We replace the architecture proposed in the original paper with CNN architecture used in \textit{sanitizer} to improve their results for higher dimensional image datasets. \underline{ii) \textit{Learned Noise:}} is built upon the TCNND architecture described in GAP~\cite{huang_context-aware_2017} where a small dimensional noise is fed to a decoder that sanitizes the image by adding the noise vector. \underline{iii) \textit{TIPRDC~\cite{li2020tiprdc}:}} is used as a baseline without any modification. \underline{iv) \textit{Noise:}} baseline sanitizes data by adding Gaussian noise in the pixel space; which is equivalent to the DP baseline used in TIPRDC~\cite{tiprdc}.
\begin{table*}[t!]
    \tabcolsep=0.45cm
    \centering
    \resizebox{0.77\textwidth}{!}{%
    \begin{tabular}{l|c|c|c|c}
    \toprule
    Method & Fairface-R $\uparrow$ & CelebA $\uparrow$ & UTKFace-R $\uparrow$ & UTKFace-G $\uparrow$ \\  \toprule
    TIPRDC~\cite{li2020tiprdc} & 0.441 & 0.465 & 0.453 & 0.443\\  \midrule
    GAP~\cite{huang_context-aware_2017} & 0.447 & 0.442 & 0.450 & 0.434\\  \midrule
    Noise & 0.438 & 0.422 & 0.435 & 0.431\\  \midrule
    Adversarial Noise~\cite{huang_context-aware_2017}* & 0.432 & 0.422 & 0.420 & 0.439 \\  \midrule
    \textbf{Ours} & \textbf{0.476} & \textbf{0.483} & \textbf{0.476} & \textbf{0.487} \\  \bottomrule
    \end{tabular}}%
    \caption{\textbf{E1: Privacy-Utility comparison:} We report area under the curve (higher is better) to compare the privacy-utility trade-offs between \textit{sanitizer} and baselines. UTKFace-R and UTKFace-G refer to the setup where \textit{race} and \textit{gender} is the sensitive attribute. Our method outperforms all baselines in all experiments.}
    \label{tab:tradeoff}
\end{table*}

\textbf{Evaluation Metrics:} We evaluate different techniques by comparing the \textit{privacy-utility trade-off}. Here the utility is measured by the data receiver's test accuracy on the downstream task using sanitized dataset. For measuring privacy, we use the technique described in Section~\ref{sec:formulation}. Specifically, we quantify information leakage from the dataset by comparing the performance of an adversary inferring sensitive information from the sanitized dataset. We simulate a strong adversary that dynamically adapts to a sanitization scheme. This adaptation is modeled by a pretrained adversary model that is finetuned on the \textit{sanitized} dataset and then evaluated on the sanitized test set. Inspired by~\cite{maxentropy} we quantify privacy-utility trade-offs curves by different techniques using area under the pareto-optimal curve 
(AuC). Higher AuC value denotes a better privacy-utility trade-off.

\subsection{Experimental Setup and Results}
\quad\ \textbf{Experiment E1} \textit{Multi-category sensitive and Binary utility}: We test the use-case \textbf{UC1} by evaluating the privacy-utility trade-off on a task where sensitive information is multi-category ``race" (fine-grained) and downstream utility task is ``gender" (coarse). Intuitively, we should get a good trade-off from all techniques that can share coarse-grained data while obfuscating fine-grained sensitive detail.

\textbf{Experiment E2} \textit{Binary sensitive and Multi-category utility}: We use the setup as E1 but use ``race" (coarse) as sensitive attribute and ``gender" (fine-grained) as utility attribute. Intuitively, we expect degradation in an overall trade-off in comparison to E1. We perform the experiments on UTKFace~\cite{utkface} dataset and call this configuration \textit{UTKFace-G}.

\textbf{Experiment E3} \textit{Single sensitive and Multiple utility}: We use the same setup as E1 but evaluate multiple utility tasks. We use CelebA~\cite{celeba} with sensitive attribute as ``gender" and utility as ``mouth open", ``smiling" and ``high cheekbone".

\textbf{Experiment E4} \textit{Learning transferable models}: We evaluate the use-case \textbf{UC2} by training a ML model on sanitized images and evaluate it on real images (non-sanitized). This setup is similar to Classification Accuracy Score(CAS) in the generative modeling community~\cite{ravuri2019classification}. Note that it is not possible to include TIPRDC baseline since their output is constrained to embedding space.

\textbf{Experiment E5} \textit{Learn sensitive attribute distribution}: We test the use-case \textbf{UC3} of learning a ML model over the distribution $p(\X,\YS)$ while protecting individual sensitive information. We train the data-receiver's ML model on $(\Xpr,\YSpr)$ and the attacker on $(\Xpr,\YS)$. We evaluate data-receiver on $(\X,\YS)$ and the attacker on $(\Xpr,\YS)$. This setup is not possible for our baselines since they censor sensitive information.

\begin{table}[!tb]
\RawFloats
    \begin{minipage}{.49\linewidth}
      \centering
        \resizebox{0.85\textwidth}{!}{%
        \begin{tabular}{l|c|c|c|c}
        \toprule
        \multicolumn{1}{l|}{} & \multicolumn{2}{c|}{UTKFace}      & \multicolumn{2}{c}{CelebA}             \\ 
        \midrule
             & Utility & Leakage & Utility & Leakage\\ 
        \midrule
            Uniform Noise          & 0.667 & 0.501 &    0.576   &  0.712      \\ \midrule
            GAP~\cite{huang_context-aware_2017}            & 0.615 & 0.499 &    0.723  &   0.686     \\ \midrule
            Adversarial Noise~\cite{huang_context-aware_2017}      & 0.801 & 0.695 &    0.746  &  \textbf{0.663}            \\ \midrule
            \textbf{Ours}              & \textbf{0.86} & \textbf{0.474} &  \textbf{0.9022} &   0.6955           \\ \bottomrule
        \end{tabular}}
        \caption{\textbf{E4, Classification Accuracy Score (CAS) evaluation}: We train a classifier on privatized data samples and evaluate them on non-privatized samples.\label{tab:E4}}
    \end{minipage}%
    \hfill
    \begin{minipage}{.49\linewidth}
      \centering
        \resizebox{0.83\textwidth}{!}{%
        \begin{tabular}{l|c|c|c|c}
        \toprule
        \multicolumn{1}{l|}{} & \multicolumn{2}{c|}{UTKFace}               & \multicolumn{2}{c}{CelebA}                                           \\ \midrule
        \multicolumn{1}{l|}{} & \multicolumn{1}{c|}{Utility} & \multicolumn{1}{c|}{Leakage} & \multicolumn{1}{c|}{Utility} & \multicolumn{1}{c}{Leakage} \\ \midrule
            Suppression            & 0.208     &        0.498   & 0.7042                             &                  0.7177                 \\ \midrule
            Obfuscation            & 0.208                             &             0.491 &               0.62               &      0.7129                                          \\ \midrule
            DP-Sampling               &            \textbf{0.521}        &            \textbf{0.474}                  &        \textbf{0.817}      &   \textbf{0.6955}                         \\ 
        \bottomrule
        \end{tabular}}
        \caption{\textbf{E5, CAS for learning a model of sensitive attribute}. We experiment with different mechanisms in the \textit{local sampling} stage.\label{tab:E5}}
    \end{minipage}
\end{table}
\textbf{Results:} For E1 and E2, we plot the privacy-utility trade-off for all techniques in Figure~\ref{fig:tradeoff} and Table~\ref{tab:tradeoff}. \textit{Sanitizer} obtains a better privacy-utility trade-off consistently. For E3, we compare trade-off by evaluating on multiple downstream tasks and observe sanitizer's consistent better performance. We posit that the consistent improvement is due to explicit modeling of different privacy-utility constraints in \textit{global-decoupler}. We compare the results for E4 in Table~\ref{tab:E4}. Unlike previous experiments, here \textit{sanitizer} achieves substantial gap in comparison to other techniques. We believe that synthetically replacing sensitive information allows \textit{sanitizer} to produce realistic sanitized samples. While the leakage is slightly larger on CelebA dataset, the relative improvement in utility is much larger. For E5, we compare three mechanisms proposed in Section~\ref{sec:mechanisms} in Table~\ref{tab:E5}. Under the same privacy budget, the proposed DP-Sampling technique achieves much better performance in both utility and leakage. Finally, we emphasize that E5 is not possible for baselines and is achieved only by the design of \textit{sanitizer} and results for E4 and E5 validate that using \textit{sanitizer} significantly improves performance for use-cases UC2 and UC3 (Section~\ref{sec:intro}).
\begin{figure}[t]
\begin{floatrow}
    \ffigbox[0.4\textwidth]{%
      \includegraphics[width=0.85\columnwidth]{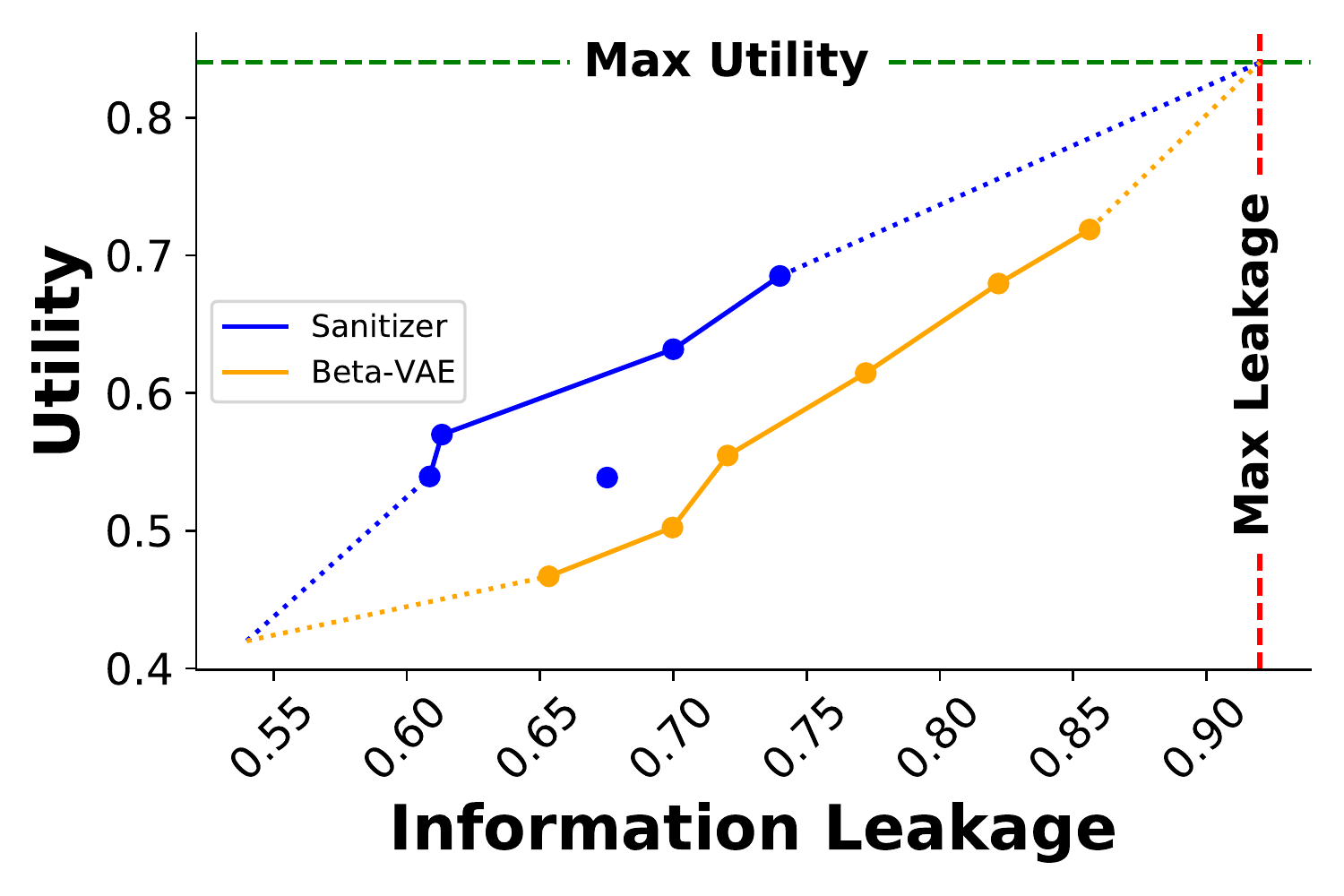}
    }{%
      \caption{\textbf{Comparing $\beta$-VAE and global-decoupler} by plotting the privacy-utility trade-off.}
      \label{fig:betavae}%
    }
    \capbtabbox[0.6\textwidth]{%
      \resizebox{0.38\textwidth}{!}{\begin{tabular}{l|c|c|c|c}
                \toprule
                Aligner & Dcorr & Adv & Leakage $\downarrow$ & Utility $\uparrow$ \\  \toprule
                \xmark & \xmark & \checkmark & 0.6259 &	0.5474 \\  \midrule
                \xmark & \checkmark & \checkmark & 0.6238 &	0.5394 \\  \midrule
                \checkmark & \xmark & \xmark & 0.6816 & 0.5137 \\  \midrule
                \checkmark & \xmark & \checkmark & 0.6318 & 0.5335 \\  \midrule
                \checkmark & \checkmark & \xmark & 0.6752 & 0.5386 \\  \midrule
                \checkmark & \checkmark & \checkmark & \textbf{0.6132} & \textbf{0.5698} \\  \midrule
      \end{tabular}}
    }{%
      \caption{\textbf{Ablation on global-decoupler} by cutting its different components. \ding{51} and \xmark\ denotes the presence and absence of the respective components.}
      \label{tab:ablation}%
    }
\end{floatrow}
\end{figure}
\subsection{Implementation Details}
All of our experiments are implemented using PyTorch~\cite{NEURIPS2019_9015} and conducted using NVIDIA 1080 Ti GPUs. We use Adam optimizer~\cite{kingma2014adam} for training all of the neural networks. Throughout all the experiments, we use a ResNet-50~\cite{resnet} architecture for prediction related tasks and transpose convolution-based architectures for generative tasks. Unless noted otherwise, we use $\epsilon=1.0$ for all the experiments. The value of $\epsilon$ is divided as $0.3$ for the mean and $0.7$ for the variance parameter of DP-sampling mechanism. For all of our evaluations we choose $k=8$ and $m=32$. Our training pipeline requires training base model for all techniques for a given set of hyper-parameters. This base model is used to obtain sanitized datasets. We use sanitized datasets to train a separate utility model and adversary model. Each utility and adversary model is trained independently for every trade-off parameter corresponding to every baseline. This results in a privacy-utility trade-off curve for each technique. We use $\alpha_1=1,\alpha_2=1,\alpha_3=100,\alpha_4=1,\beta=5$ for all sanitizer experiments except the ones where these parameters were changed to obtain a trade-off. We keep $\alpha_3=100$ since the distance correlation is typically a small quantity and requires scaling in order to influence the overall loss term.\\
\subsection{Datasets}
\textbf{UTKFace~\cite{utkface}} consists of 20,000 face images. We use the cropped and aligned version of the dataset and generate a random split of $90\%-10\%$, training and testing data. The dataset has ``ethnicity", ``gender", and ``age" as categorical labels. For our experiments, we keep the sensitive attribute as ethnicity which has 5 unique labels and due to class imbalance, the best possible performance without access to the image is 44\%. We use ``gender" as the utility attribute for the evaluation.\\
\textbf{CelebA~\cite{celeba}} is a large scale dataset of 202,599 celebrity face images 10,177 unique identities, each with 40 binary attribute annotations. For our experiments, we define gender as the sensitive attribute. We use ``mouth open", ``smiling" and ``high cheekbones" as the utility attribute for evaluation evaluation.\\
\textbf{FairFace~\cite{fairface}} dataset consists of 108,501 images, with three different attributes ``ethnicity", ``gender", and ``age". The dataset was curated to overcome the class imbalance in ethnicity present in the majority of the existing Faces datasets. We use ``ethnicity" as a sensitive attribute. We use ``gender" as the attribute for the utility evaluation.
\section{Discussion}
\label{sec:discussion}
Here, we analyze the design of global-decoupler by performing ablation study and discuss architectural limitations associated with it.

\underline{\textbf{i) Ablation study for global-decoupler:}}
We perform ablation on each of the components described in the architecture in Figure~\ref{fig:arch}. We measure the sensitive information leakage and utility by comparing performance with and without each component in the objective given in Section~\ref{sec:method}. We can interpret this ablation study as keeping $\alpha_i=0$ for the $i$'th component during the \textit{global decoupling} stage. We enumerate the results in Table~\ref{tab:ablation}. We note that the presence of all components provides the best trade-off between the leakage and utility.\\
The global-decoupler is built upon $\beta$-VAE; therefore, we compare the trade-off between the two. We utilize latent space interpolation to randomize the sensitive attribute. First, we train a $\beta$-VAE model and obtain the mean representation of the sensitive latent by $z_{S_i} = \frac{1}{n_i}\sum_{j\in S_i}z_j~\sim q_\phi(z|x_j)$ where $S_i$ refers to a unique sensitive attribute category with $n_i$ number of samples in the dataset. Finally, to randomize sensitive information in a given sample $x$, we transform the original latent $z = q_\phi(x)$ to obtain a sanitized latent $\tilde{z} = z - z_{S_i} + z_{S_j}$. For performing sensitive attribute randomization, $i$ is the sensitive category of $z$, and $j$ is chosen uniformly from the set of all categories including $i$. Finally, we obtain $\xpr = p_\theta(\tilde{z})$ as a sanitized transformation of $x$. We evaluate this technique on UTKFace dataset~\cite{utkface} with same experimental setup as E2. We model the attacker same way as explained in Section~\ref{sec:exp} and show trade-off curves in Figure~\ref{fig:betavae}.

\underline{\textbf{ii) Architectural Limitations: }} The key goal of this work is to introduce a systematic framework and mechanisms for sanitization that could be useful for as many downstream tasks as possible under the privacy-utility trade-off. Here, we note two key limitations of the presented results, emerging from the generative modeling framework: i) \textit{input sample size} - This limitation stems from the need for sufficient data points to learn a latent model of the data that generalizes between $\Daux$ and $\DA$. Designing latent models that can capture the distribution with a minimum number of samples is an active area of research in few-shot learning which will improve the impact of our results but is orthogonal to the scope of our work.
ii) \textit{output sample quality} - This limitation can be improved using hierarchical latent variable models~\cite{vahdat2020nvae,razavi2019generating} and we consider this as part of the future work. We believe that improvement in representation capacity can improve trade-offs even further for \textit{sanitizer}.
\section{Related Work}
\textit{First}, we discuss prior work in privacy-preserving data release for task-dependent and task-independent setups. \textit{Next}, we draw parallels to techniques in fairness and conditional generation of images.

\textbf{Task-dependent} data release techniques transform data that is conducive to a particular task. Several techniques use central DP~\cite{dwork2014algorithmic} as a formal privacy definition to answer aggregate queries. The queries can be summary statistics such as mean, median~\cite{dwork2006calibrating,dwork2014algorithmic} or learning a ML model~\cite{dpsgd}, sharing gradients in federated learning~\cite{wei2020federated}. In addition to being task-dependent, the central DP paradigm is different from the sanitization problem in two other ways: i) Central DP minimizes leakage of every identifiable information while \textit{sanitizer} minimizes leakage of only presupposed secret information, ii) Central DP performs a query on dataset aggregated by a trusted curator before sharing it with untrusted parties while \textit{sanitizer} does not have the notion of a trusted curator and hence, sanitization is performed by the data owner.
Recent work in adversarial learning has resulted in techniques for task-specific latent representation~\cite{deeprotect,wu2018towards,JMLR:v18:16-501,Roy_2019_CVPR,li2019deepobfuscator,osia2018deep,osia_hybrid_2020,anonymous_private_2020,shredder,singh2020disco}. While we share the same goal of protecting sensitive information, our work differs in its task-independent formulation.

\textbf{Task-independent} techniques share data in a non-interactive manner. Similar to central DP, several works consider identification as sensitive information, however without a trusted curator. This modified central DP setup is referred to as local-DP~\cite{raskhodnikova2008can}. While variants of local-DP for attribute privacy exist, their focus is primarily on protecting dataset statistics~\cite{zhang2018mitigating}, different rows of a dataset~\cite{acharya2020context} or task-dependent~\cite{cheng2021task,murakami2019utility}. Local-DP based generative models~\cite{jordon_pate-gan_2018,xie2018differentially,torkzadehmahani2019dp,chanyaswad2019ron,zhang2021privsyn} learn data distribution privately to release samples. While we focus on specific sensitive information, we build upon the sampling strategy used in RonGauss~\cite{chanyaswad2019ron} to sample sensitive data. TIPRDC~\cite{tiprdc} and GAP~\cite{gap} are task-agnostic techniques that protect sensitive information by censoring it. While we solve a similar problem, our sampling-based approach allows performing certain tasks (eg. \textbf{E5} in Section~\ref{sec:exp}) that are not possible with the censoring-based approach.

\textbf{Fairness} techniques aim to make predictive models unbiased with respect to \textit{protected groups}. Among different approaches~\cite{caton2020fairness} used for fairness, works in censoring information~\cite{zemel2013learning,wang2019balanced,adeli2019bias,zhang2018mitigating} related to \textit{protected groups} is the closest approach to our work. However, we differ significantly from the censoring approach because we release anonymized sensitive information instead of censoring it. Furthermore, the goal of the sanitization problem is to maximally retain original data insofar that all biases would exist after sanitization. While the objective and evaluation for the fairness community are different, we note that Sarhan et al.~\cite{sarhan2020fairness} use a similar objective as \textit{sanitizer} by utilizing variational inference with orthogonality constraint for preventing leakage. However, they do not provide anonymization since two correlated vectors can be orthogonal.

\textbf{Conditional generation} which has a similar problem setup to sanitization~\cite{semantic1,semantic2}. While this has led to some relevant work in privacy, the techniques typically handcraft the objective to be task-specific for identity~\cite{chhabra2018anonymizing,mirjalili2020privacynet,mirjalili2019flowsan,othman2014privacy,raynal2020image}. In contrast, \textit{sanitizer} is agnostic of target utility and only depends upon sensitive attributes. Some recent works utilize uncertainty-based metrics~\cite{wang2021infoscrub,martinsson2020adversarial} to fuse all sensitive attributes in latent space using adversarial training but hence, unlike \textit{sanitizer}, generate highly unrealistic images (hence not being task agnostic) due to high uncertainty.
\vspace{-1mm}
\section{Conclusion}
In this work, we presented sanitizer: a framework for minimizing sensitive information leakage to facilitate task-agnostic data release. To the best of our knowledge, this is the first work to show how to hide sensitive information in data while enabling the data receiver to learn the distribution of sensitive information. We achieve this objective through a two-stage process - i) \textit{global decoupling} for learning a latent model of data and ii) \textit{local sampling} for securely synthesizing sensitive information. While our approach demonstrates a relatively good privacy-utility trade-off, it is possible to further decrease the sensitive information leakage in the sanitized version by improving regularization and the training method used in this paper. Future work includes using \textit{sanitizer} for task-dependent data release and extension to other modalities where unstructured data is involved such as speech, NLP, and time-series data.

\bibliographystyle{plain}
\bibliography{main}

\begin{thebibliography}{10}

\bibitem{dpsgd}
Martin Abadi, Andy Chu, Ian Goodfellow, H.~Brendan McMahan, Ilya Mironov, Kunal
  Talwar, and Li~Zhang.
\newblock Deep learning with differential privacy.
\newblock {\em Proceedings of the 2016 ACM SIGSAC Conference on Computer and
  Communications Security}, Oct 2016.

\bibitem{acharya2020context}
Jayadev Acharya, Kallista Bonawitz, Peter Kairouz, Daniel Ramage, and Ziteng
  Sun.
\newblock Context aware local differential privacy.
\newblock In {\em International Conference on Machine Learning}, pages 52--62.
  PMLR, 2020.

\bibitem{adeli2019bias}
Ehsan Adeli, Qingyu Zhao, Adolf Pfefferbaum, Edith~V Sullivan, Li~Fei-Fei,
  Juan~Carlos Niebles, and Kilian~M Pohl.
\newblock Bias-resilient neural network.
\newblock 2019.

\bibitem{banerjee2021reading}
Imon Banerjee, Ananth~Reddy Bhimireddy, John~L Burns, Leo~Anthony Celi,
  Li-Ching Chen, Ramon Correa, Natalie Dullerud, Marzyeh Ghassemi, Shih-Cheng
  Huang, Po-Chih Kuo, et~al.
\newblock Reading race: Ai recognises patient's racial identity in medical
  images.
\newblock {\em arXiv preprint arXiv:2107.10356}, 2021.

\bibitem{betzler2021gender}
Bjorn~Kaijun Betzler, Henrik Hee~Seung Yang, Sahil Thakur, Marco Yu, Zhi
  Da~Soh, Geunyoung Lee, Yih-Chung Tham, Tien~Yin Wong, Tyler~Hyungtaek Rim,
  Ching-Yu Cheng, et~al.
\newblock Gender prediction for a multiethnic population via deep learning
  across different retinal fundus photograph fields: Retrospective
  cross-sectional study.
\newblock {\em JMIR medical informatics}, 9(8):e25165, 2021.

\bibitem{borgwardt2006integrating}
Karsten~M Borgwardt, Arthur Gretton, Malte~J Rasch, Hans-Peter Kriegel,
  Bernhard Sch{\"o}lkopf, and Alex~J Smola.
\newblock Integrating structured biological data by kernel maximum mean
  discrepancy.
\newblock {\em Bioinformatics}, 22(14):e49--e57, 2006.

\bibitem{caton2020fairness}
Simon Caton and Christian Haas.
\newblock Fairness in machine learning: A survey.
\newblock {\em arXiv preprint arXiv:2010.04053}, 2020.

\bibitem{chanyaswad2019ron}
Thee Chanyaswad, Changchang Liu, and Prateek Mittal.
\newblock Ron-gauss: Enhancing utility in non-interactive private data release.
\newblock {\em Proceedings on Privacy Enhancing Technologies}, 2019(1):26--46,
  2019.

\bibitem{tcvae}
Ricky~TQ Chen, Xuechen Li, Roger Grosse, and David Duvenaud.
\newblock Isolating sources of disentanglement in variational autoencoders.
\newblock {\em arXiv:1802.04942}, 2018.

\bibitem{chen2018development}
Shi Chen, Zhou-xian Pan, Hui-juan Zhu, Qing Wang, Ji-Jiang Yang, Yi~Lei,
  Jian-qiang Li, and Hui Pan.
\newblock Development of a computer-aided tool for the pattern recognition of
  facial features in diagnosing turner syndrome: comparison of diagnostic
  accuracy with clinical workers.
\newblock {\em Scientific reports}, 8(1):1--9, 2018.

\bibitem{semantic2}
Ying-Cong Chen, Xiaohui Shen, Zhe Lin, Xin Lu, I~Pao, Jiaya Jia, et~al.
\newblock Semantic component decomposition for face attribute manipulation.
\newblock In {\em CVPR}, 2019.

\bibitem{cheng2021task}
Jiangnan Cheng, Ao~Tang, and Sandeep Chinchali.
\newblock Task-aware privacy preservation for multi-dimensional data.
\newblock {\em arXiv preprint arXiv:2110.02329}, 2021.

\bibitem{chhabra2018anonymizing}
Saheb Chhabra, Richa Singh, Mayank Vatsa, and Gaurav Gupta.
\newblock Anonymizing k-facial attributes via adversarial perturbations.
\newblock {\em arXiv preprint arXiv:1805.09380}, 2018.

\bibitem{deng2009imagenet}
Jia Deng, Wei Dong, Richard Socher, Li-Jia Li, Kai Li, and Li~Fei-Fei.
\newblock Imagenet: A large-scale hierarchical image database.
\newblock In {\em 2009 IEEE conference on computer vision and pattern
  recognition}, pages 248--255. Ieee, 2009.

\bibitem{dwork_calibrating_2006}
Cynthia Dwork, Frank McSherry, Kobbi Nissim, and Adam Smith.
\newblock Calibrating {Noise} to {Sensitivity} in {Private} {Data} {Analysis}.
\newblock In Shai Halevi and Tal Rabin, editors, {\em Theory of
  {Cryptography}}, Lecture {Notes} in {Computer} {Science}, pages 265--284,
  Berlin, Heidelberg, 2006. Springer.

\bibitem{dwork2006calibrating}
Cynthia Dwork, Frank McSherry, Kobbi Nissim, and Adam Smith.
\newblock Calibrating noise to sensitivity in private data analysis.
\newblock In {\em Theory of cryptography conference}, 2006.

\bibitem{dwork2014algorithmic}
Cynthia Dwork, Aaron Roth, et~al.
\newblock The algorithmic foundations of differential privacy.
\newblock {\em Foundations and Trends in Theoretical Computer Science},
  9(3-4):211--407, 2014.

\bibitem{goodfellow2014generative}
Ian~J Goodfellow, Jean Pouget-Abadie, Mehdi Mirza, Bing Xu, David Warde-Farley,
  Sherjil Ozair, Aaron Courville, and Yoshua Bengio.
\newblock Generative adversarial networks.
\newblock {\em arXiv:1406.2661}, 2014.

\bibitem{gretton2005measuring}
Arthur Gretton, Olivier Bousquet, Alex Smola, and Bernhard Sch{\"o}lkopf.
\newblock Measuring statistical dependence with hilbert-schmidt norms.
\newblock In {\em International conference on algorithmic learning theory},
  pages 63--77. Springer, 2005.

\bibitem{JMLR:v18:16-501}
Jihun Hamm.
\newblock Minimax filter: Learning to preserve privacy from inference attacks.
\newblock {\em Journal of Machine Learning Research}, 18(129):1--31, 2017.

\bibitem{resnet}
Kaiming He, Xiangyu Zhang, Shaoqing Ren, and Jian Sun.
\newblock Deep residual learning for image recognition.
\newblock {\em CoRR}, abs/1512.03385, 2015.

\bibitem{betavae}
Irina Higgins, Loic Matthey, Arka Pal, Christopher Burgess, Xavier Glorot,
  Matthew Botvinick, Shakir Mohamed, and Alexander Lerchner.
\newblock beta-vae: Learning basic visual concepts with a constrained
  variational framework.
\newblock 2016.

\bibitem{huang_context-aware_2017}
Chong Huang, Peter Kairouz, Xiao Chen, Lalitha Sankar, and Ram Rajagopal.
\newblock Context-{Aware} {Generative} {Adversarial} {Privacy}.
\newblock {\em Entropy}, 19(12):656, December 2017.
\newblock arXiv: 1710.09549.

\bibitem{gap}
Chong Huang, Peter Kairouz, Xiao Chen, Lalitha Sankar, and Ram Rajagopal.
\newblock Generative adversarial privacy.
\newblock {\em CoRR}, 2018.

\bibitem{jordon_pate-gan_2018}
James Jordon, Jinsung Yoon, and Mihaela van~der Schaar.
\newblock {PATE}-{GAN}: {Generating} {Synthetic} {Data} with {Differential}
  {Privacy} {Guarantees}.
\newblock September 2018.

\bibitem{fairface}
Kimmo K{\"a}rkk{\"a}inen and Jungseock Joo.
\newblock Fairface: Face attribute dataset for balanced race, gender, and age.
\newblock {\em arXiv:1908.04913}, 2019.

\bibitem{kifer2014pufferfish}
Daniel Kifer and Ashwin Machanavajjhala.
\newblock Pufferfish: A framework for mathematical privacy definitions.
\newblock {\em ACM TODS}, 2014.

\bibitem{factorvae}
Hyunjik Kim and Andriy Mnih.
\newblock Disentangling by factorising.
\newblock In {\em ICML}, 2018.

\bibitem{kingma2014adam}
Diederik~P Kingma and Jimmy Ba.
\newblock Adam: A method for stochastic optimization.
\newblock {\em arXiv:1412.6980}, 2014.

\bibitem{vae}
Diederik~P Kingma and Max Welling.
\newblock Auto-encoding variational bayes.
\newblock {\em arXiv:1312.6114}, 2013.

\bibitem{korot2021predicting}
Edward Korot, Nikolas Pontikos, Xiaoxuan Liu, Siegfried~K Wagner, Livia Faes,
  Josef Huemer, Konstantinos Balaskas, Alastair~K Denniston, Anthony Khawaja,
  and Pearse~A Keane.
\newblock Predicting sex from retinal fundus photographs using automated deep
  learning.
\newblock {\em Scientific reports}, 11(1):1--8, 2021.

\bibitem{kumar2021cardiac}
Deepak Kumar, Chaman Verma, Sanjay Dahiya, Pradeep~Kumar Singh, and
  Maria~Simona Raboaca.
\newblock Cardiac diagnostic feature and demographic identification models: A
  futuristic approach for smart healthcare using machine learning.
\newblock 2021.

\bibitem{tiprdc}
Ang Li, Yixiao Duan, Huanrui Yang, Yiran Chen, and Jianlei Yang.
\newblock Tiprdc: Task-independent privacy-respecting data crowdsourcing
  framework for deep learning with anonymized intermediate representations.
\newblock In {\em ACM SIGKDD}, 2020.

\bibitem{li2020tiprdc}
Ang Li, Yixiao Duan, Huanrui Yang, Yiran Chen, and Jianlei Yang.
\newblock Tiprdc: task-independent privacy-respecting data crowdsourcing
  framework for deep learning with anonymized intermediate representations.
\newblock In {\em Proceedings of the 26th ACM SIGKDD International Conference
  on Knowledge Discovery \& Data Mining}, pages 824--832, 2020.

\bibitem{li2019deepobfuscator}
Ang Li, Jiayi Guo, Huanrui Yang, and Yiran Chen.
\newblock Deepobfuscator: Adversarial training framework for privacy-preserving
  image classification, 2019.

\bibitem{deeprotect}
Changchang Liu, Supriyo Chakraborty, and Prateek Mittal.
\newblock Deeprotect: Enabling inference-based access control on mobile sensing
  applications.
\newblock {\em CoRR}, 2017.

\bibitem{celeba}
Ziwei Liu, Ping Luo, Xiaogang Wang, and Xiaoou Tang.
\newblock Large-scale celebfaces attributes (celeba) dataset.
\newblock {\em Retrieved August}, 2018.

\bibitem{loos2003computer}
Hartmut~S Loos, Dagmar Wieczorek, Rolf~P W{\"u}rtz, Christoph von~der Malsburg,
  and Bernhard Horsthemke.
\newblock Computer-based recognition of dysmorphic faces.
\newblock {\em European Journal of Human Genetics}, 11(8):555--560, 2003.

\bibitem{makhdoumi2013privacy}
Ali Makhdoumi and Nadia Fawaz.
\newblock Privacy-utility tradeoff under statistical uncertainty.
\newblock In {\em Annual Allerton Conference on Communication, Control, and
  Computing (Allerton)}, 2013.

\bibitem{martinsson2020adversarial}
John Martinsson, Edvin~Listo Zec, Daniel Gillblad, and Olof Mogren.
\newblock Adversarial representation learning for synthetic replacement of
  private attributes.
\newblock {\em arXiv preprint arXiv:2006.08039}, 2020.

\bibitem{mclachlan2019finite}
Geoffrey~J McLachlan, Sharon~X Lee, and Suren~I Rathnayake.
\newblock Finite mixture models.
\newblock {\em Annual review of statistics and its application}, 6:355--378,
  2019.

\bibitem{shredder}
Fatemehsadat Mireshghallah, Mohammadkazem Taram, Prakash Ramrakhyani, Dean~M.
  Tullsen, and Hadi Esmaeilzadeh.
\newblock Shredder: Learning noise to protect privacy with partial {DNN}
  inference on the edge.
\newblock {\em CoRR}, abs/1905.11814, 2019.

\bibitem{mirjalili2019flowsan}
Vahid Mirjalili, Sebastian Raschka, and Arun Ross.
\newblock Flowsan: Privacy-enhancing semi-adversarial networks to confound
  arbitrary face-based gender classifiers.
\newblock {\em IEEE Access}, 7:99735--99745, 2019.

\bibitem{mirjalili2020privacynet}
Vahid Mirjalili, Sebastian Raschka, and Arun Ross.
\newblock Privacynet: semi-adversarial networks for multi-attribute face
  privacy.
\newblock {\em IEEE Transactions on Image Processing}, 29:9400--9412, 2020.

\bibitem{murakami2019utility}
Takao Murakami and Yusuke Kawamoto.
\newblock $\{$Utility-Optimized$\}$ local differential privacy mechanisms for
  distribution estimation.
\newblock In {\em 28th USENIX Security Symposium (USENIX Security 19)}, pages
  1877--1894, 2019.

\bibitem{osia2020hybrid}
Seyed~Ali Osia, Ali~Shahin Shamsabadi, Sina Sajadmanesh, Ali Taheri, Kleomenis
  Katevas, Hamid~R Rabiee, Nicholas~D Lane, and Hamed Haddadi.
\newblock A hybrid deep learning architecture for privacy-preserving mobile
  analytics.
\newblock {\em IEEE Internet of Things Journal}, 2020.

\bibitem{osia_hybrid_2020}
Seyed~Ali Osia, Ali~Shahin Shamsabadi, Sina Sajadmanesh, Ali Taheri, Kleomenis
  Katevas, Hamid~R. Rabiee, Nicholas~D. Lane, and Hamed Haddadi.
\newblock A {Hybrid} {Deep} {Learning} {Architecture} for
  {Privacy}-{Preserving} {Mobile} {Analytics}.
\newblock {\em IEEE Internet of Things Journal}, 7(5):4505--4518, May 2020.
\newblock arXiv: 1703.02952.

\bibitem{osia2018deep}
Seyed~Ali Osia, Ali Taheri, Ali~Shahin Shamsabadi, Kleomenis Katevas, Hamed
  Haddadi, and Hamid~R. Rabiee.
\newblock Deep private-feature extraction, 2018.

\bibitem{othman2014privacy}
Asem Othman and Arun Ross.
\newblock Privacy of facial soft biometrics: Suppressing gender but retaining
  identity.
\newblock In {\em European Conference on Computer Vision}, pages 682--696.
  Springer, 2014.

\bibitem{NEURIPS2019_9015}
Adam Paszke et~al.
\newblock Pytorch: An imperative style, high-performance deep learning library.
\newblock In {\em NeurIPS}. 2019.

\bibitem{raskhodnikova2008can}
Sofya Raskhodnikova, Adam Smith, Homin~K Lee, Kobbi Nissim, and Shiva~Prasad
  Kasiviswanathan.
\newblock What can we learn privately.
\newblock In {\em FOCS}, 2008.

\bibitem{ravuri2019classification}
Suman Ravuri and Oriol Vinyals.
\newblock Classification accuracy score for conditional generative models.
\newblock {\em arXiv:1905.10887}, 2019.

\bibitem{raynal2020image}
Mathilde Raynal, Radhakrishna Achanta, and Mathias Humbert.
\newblock Image obfuscation for privacy-preserving machine learning.
\newblock {\em arXiv preprint arXiv:2010.10139}, 2020.

\bibitem{razavi2019generating}
Ali Razavi, Aaron Van~den Oord, and Oriol Vinyals.
\newblock Generating diverse high-fidelity images with vq-vae-2.
\newblock {\em Advances in neural information processing systems}, 32, 2019.

\bibitem{rezende2015variational}
Danilo Rezende and Shakir Mohamed.
\newblock Variational inference with normalizing flows.
\newblock In {\em International conference on machine learning}, pages
  1530--1538. PMLR, 2015.

\bibitem{rezende2014stochastic}
Danilo~Jimenez Rezende, Shakir Mohamed, and Daan Wierstra.
\newblock Stochastic backpropagation and approximate inference in deep
  generative models.
\newblock In {\em ICML}, 2014.

\bibitem{maxentropy}
Proteek~Chandan Roy and Vishnu~Naresh Boddeti.
\newblock Mitigating information leakage in image representations: A maximum
  entropy approach.
\newblock In {\em CVPR}, 2019.

\bibitem{Roy_2019_CVPR}
Proteek~Chandan Roy and Vishnu~Naresh Boddeti.
\newblock Mitigating information leakage in image representations: A maximum
  entropy approach.
\newblock In {\em Proceedings of the IEEE/CVF Conference on Computer Vision and
  Pattern Recognition (CVPR)}, June 2019.

\bibitem{anonymous_private_2020}
Mohammad Samragh, Hossein Hosseini, Aleksei Triastcyn, Kambiz Azarian, Joseph
  Soriaga, and Farinaz Koushanfar.
\newblock Unsupervised information obfuscation for split inference of neural
  networks.
\newblock {\em arXiv preprint arXiv:2104.11413}, 2021.

\bibitem{sankar2010information}
Lalitha Sankar, S~Raj Rajagopalan, and H~Vincent Poor.
\newblock An information-theoretic approach to privacy.
\newblock In {\em Allerton Conference on Communication, Control, and Computing
  (Allerton)}, 2010.

\bibitem{sarhan2020fairness}
Mhd~Hasan Sarhan, Nassir Navab, Abouzar Eslami, and Shadi Albarqouni.
\newblock Fairness by learning orthogonal disentangled representations.
\newblock In {\em European Conference on Computer Vision}, pages 746--761.
  Springer, 2020.

\bibitem{semantic1}
Wei Shen and Rujie Liu.
\newblock Learning residual images for face attribute manipulation.
\newblock In {\em CVPR}, 2017.

\bibitem{singh2020disco}
Abhishek Singh, Ayush Chopra, Vivek Sharma, Ethan Garza, Emily Zhang, Praneeth
  Vepakomma, and Ramesh Raskar.
\newblock Disco: Dynamic and invariant sensitive channel obfuscation for deep
  neural networks.
\newblock {\em arXiv:2012.11025}, 2020.

\bibitem{stephen2017facial}
Ian~D Stephen, Vivian Hiew, Vinet Coetzee, Bernard~P Tiddeman, and David~I
  Perrett.
\newblock Facial shape analysis identifies valid cues to aspects of
  physiological health in caucasian, asian, and african populations.
\newblock {\em Frontiers in psychology}, 8:1883, 2017.

\bibitem{su2021affective}
Norman~Makoto Su and David~J Crandall.
\newblock The affective growth of computer vision.
\newblock In {\em Proceedings of the IEEE/CVF Conference on Computer Vision and
  Pattern Recognition}, pages 9291--9300, 2021.

\bibitem{szekely2007measuring}
G{\'a}bor~J Sz{\'e}kely, Maria~L Rizzo, Nail~K Bakirov, et~al.
\newblock Measuring and testing dependence by correlation of distances.
\newblock {\em The annals of statistics}, 2007.

\bibitem{tao2021benchmarking}
Yuchao Tao, Ryan McKenna, Michael Hay, Ashwin Machanavajjhala, and Gerome
  Miklau.
\newblock Benchmarking differentially private synthetic data generation
  algorithms.
\newblock {\em arXiv preprint arXiv:2112.09238}, 2021.

\bibitem{torkzadehmahani2019dp}
Reihaneh Torkzadehmahani, Peter Kairouz, and Benedict Paten.
\newblock Dp-cgan: Differentially private synthetic data and label generation.
\newblock In {\em Proceedings of the IEEE/CVF Conference on Computer Vision and
  Pattern Recognition Workshops}, pages 0--0, 2019.

\bibitem{vahdat2020nvae}
Arash Vahdat and Jan Kautz.
\newblock Nvae: A deep hierarchical variational autoencoder.
\newblock {\em Advances in Neural Information Processing Systems},
  33:19667--19679, 2020.

\bibitem{vepakomma2020nopeek}
Praneeth Vepakomma, Abhishek Singh, Emily Zhang, Otkrist Gupta, and Ramesh
  Raskar.
\newblock Nopeek-infer: Preventing face reconstruction attacks in distributed
  inference after on-premise training.
\newblock In {\em 2021 16th IEEE International Conference on Automatic Face and
  Gesture Recognition (FG 2021)}, pages 1--8. IEEE, 2021.

\bibitem{wang2021infoscrub}
Hui-Po Wang, Tribhuvanesh Orekondy, and Mario Fritz.
\newblock Infoscrub: Towards attribute privacy by targeted obfuscation.
\newblock In {\em Proceedings of the IEEE/CVF Conference on Computer Vision and
  Pattern Recognition}, pages 3281--3289, 2021.

\bibitem{wang2019balanced}
Tianlu Wang, Jieyu Zhao, Mark Yatskar, Kai-Wei Chang, and Vicente Ordonez.
\newblock Balanced datasets are not enough: Estimating and mitigating gender
  bias in deep image representations.
\newblock In {\em Proceedings of the IEEE/CVF International Conference on
  Computer Vision}, pages 5310--5319, 2019.

\bibitem{wei2020federated}
Kang Wei, Jun Li, Ming Ding, Chuan Ma, Howard~H Yang, Farhad Farokhi, Shi Jin,
  Tony~QS Quek, and H~Vincent Poor.
\newblock Federated learning with differential privacy: Algorithms and
  performance analysis.
\newblock {\em IEEE Transactions on Information Forensics and Security},
  15:3454--3469, 2020.

\bibitem{wu2018towards}
Zhenyu Wu, Zhangyang Wang, Zhaowen Wang, and Hailin Jin.
\newblock Towards privacy-preserving visual recognition via adversarial
  training: A pilot study.
\newblock In {\em Proceedings of the European Conference on Computer Vision
  (ECCV)}, pages 606--624, 2018.

\bibitem{xie2018differentially}
Liyang Xie, Kaixiang Lin, Shu Wang, Fei Wang, and Jiayu Zhou.
\newblock Differentially private generative adversarial network.
\newblock {\em arXiv preprint arXiv:1802.06739}, 2018.

\bibitem{yi2021radiology}
Paul~H Yi, Jinchi Wei, Tae~Kyung Kim, Jiwon Shin, Haris~I Sair, Ferdinand~K
  Hui, Gregory~D Hager, and Cheng~Ting Lin.
\newblock Radiology “forensics”: determination of age and sex from chest
  radiographs using deep learning.
\newblock {\em Emergency Radiology}, 28(5):949--954, 2021.

\bibitem{zemel2013learning}
Rich Zemel, Yu~Wu, Kevin Swersky, Toni Pitassi, and Cynthia Dwork.
\newblock Learning fair representations.
\newblock In {\em International conference on machine learning}, pages
  325--333. PMLR, 2013.

\bibitem{zhang2018mitigating}
Brian~Hu Zhang, Blake Lemoine, and Margaret Mitchell.
\newblock Mitigating unwanted biases with adversarial learning.
\newblock In {\em Proceedings of the 2018 AAAI/ACM Conference on AI, Ethics,
  and Society}, pages 335--340, 2018.

\bibitem{utkface}
Zhifei Zhang, Yang Song, and Hairong Qi.
\newblock Age progression/regression by conditional adversarial autoencoder.
\newblock In {\em Proceedings of the IEEE conference on computer vision and
  pattern recognition}, pages 5810--5818, 2017.

\bibitem{zhang2021privsyn}
Zhikun Zhang, Tianhao Wang, Jean Honorio, Ninghui Li, Michael Backes, Shibo He,
  Jiming Chen, and Yang Zhang.
\newblock Privsyn: Differentially private data synthesis.
\newblock 2021.

\end{thebibliography}

\end{document}